# Targeting predictors in random forest regression[*]

Daniel Borup[†]  Bent Jesper Christensen[‡]

Nicolaj Søndergaard Mühlbach[§]  Mikkel Slot Nielsen[¶]

**Abstract**

Random forest regression (RF) is an extremely popular tool for the analysis of high-dimensional data. Nonetheless, its benefits may be lessened in sparse settings due to weak predictors, and a pre-estimation dimension reduction (targeting) step is required. We show that proper targeting controls the probability of placing splits along strong predictors, thus providing an important complement to RF's feature sampling. This is supported by simulations using representative finite samples. Moreover, we quantify the immediate gain from targeting in terms of increased strength of individual trees. Macroeconomic and financial applications show that the bias-variance trade-off implied by targeting, due to increased correlation among trees in the forest, is balanced at a medium degree of targeting, selecting the best 10–30% of commonly applied predictors. Improvements in predictive accuracy of targeted RF relative to ordinary RF are considerable, up to 12–13%, occurring both in recessions and expansions, particularly at long horizons.

[*]We are grateful to Guido W. Imbens, Stefan Wager, Christian B. Hansen, Bo Honoré, Jan Pedersen, Erik Christian Montes Schütte, and participants at the Econometric Society/Bocconi University World Congress (2020), 2nd Vienna Workshop on Economic Forecasting (2020), and Economics seminar (2020) at Aarhus University for useful discussions and comments, and to the Center for Research in Econometric Analysis of TimE Series (CREATES), the Dale T. Mortensen Center, Aarhus University, and the Danish Council for Independent Research (Grant 9056 - 00011B) for research support. Bent Jesper Christensen and Daniel Borup are also Danish Finance Institute (DFI) research fellows.

[†]Corresponding author. CREATES, Department of Economics and Business Economics, Aarhus University, Fuglesangs Allé 4, 8210 Aarhus V, Denmark, and the Danish Finance Institute (DFI). Email: dborup@econ.au.dk.

[‡]CREATES and the Dale T. Mortensen Center, Department of Economics and Business Economics, Aarhus University, and the Danish Finance Institute (DFI). Email: bjchristensen@econ.au.dk.

[§]Department of Economics, Massachusetts Institute of Technology, and CREATES. Email: muhlbach@mit.edu.

[¶]Department of Statistics, Columbia University, and CREATES. Email: m.nielsen@columbia.edu.

# 1. Introduction

Recent trends in economic forecasting have emphasized the use of machine learning techniques in settings with many predictors. The method of random forest (RF) regression (Breiman, 2001; Amit and Geman, 1997; Ho, 1998) is particularly popular due to its wide applicability, allowance for nonlinearity in data, and adaptability to high-dimensional feature spaces (many predictors) among other things. According to Howard and Bowles (2012), RF has been the most successful general-purpose algorithm in modern times. It is best described as a "divide and conquer" approach that bootstraps fractions of data, grows a decision tree on each fraction, and then aggregates these predictions.

Tree diversity is ensured by the bootstrap step and feature sampling which, at each node in the tree, restricts the possible split directions to a randomly chosen subset of the predictors (Wager, 2016). RF is easy to apply and is implemented in most programming languages. For instance, it can be found in the **sklearn** library in Python, the **randomForest** and **ranger** packages in R, and the **TreeBagger** class in MATLAB. Several fields within economics and finance have adopted tree-based algorithms as a data-driven approach to inference and forecasting, e.g., Athey et al. (2019) recast a classical kernel weighting function as an adaptive weighting function based on RF, Wager and Athey (2018) estimate heterogeneous treatment effects, Ng (2014) employs trees to forecast economic recessions, and Gu et al. (2020) use RF to predict future stock returns using numerous firm-specific and common predictors.

Although it is widely acknowledged that RF is applicable in high-dimensional settings as it has the potential to detect informative predictors automatically (see, e.g., Biau and Scornet, 2016), the need to select a reduced number of predictors from the full list of initial candidates, prior to implementing a particular forecasting method, has been emphasized in recent literature. According to Gentzkow et al. (2019):

> "The benefits of regression trees—nonlinearity and high-order interactions—are sometimes lessened in the presence of high-dimensional inputs. […]. Often times, a more beneficial use of trees is in a final prediction step after some dimension reduction […]."

Thus, RF applied in high dimensions without an initial weeding out of irrelevant predictors may fail to reach its full potential.

This principle of *targeting predictors* in high-dimensional settings, i.e., an initial (supervised) dimension reduction step before feeding data into an algorithm, was introduced by Bai and Ng (2008) in the context of factor-based prediction. Targeting is typically achieved via regularization, such as the LASSO (Least Absolute Shrinkage and Selection Operator) of Tibshirani (1996), or a related method, e.g., elastic net (Zou and Hastie, 2005), adaptive LASSO (Zou, 2006), or Bayesian shrinkage (De Mol et al., 2008). It involves choosing the number of predictors to target, which therefore effectively constitutes a tuning parameter of the procedure. Further, screening techniques may be applied, see, e.g., Fan and Lv (2008), Fan et al. (2011), and the review by Fan and Lv (2010) of variable selection in high dimensions, including both regularization and screening. In the paper, we take targeting as given to focus on its effects on RF. Thus, we do not compare different targeting techniques.

Targeting of predictors has recently been applied in various high-dimensional prediction problems. Kotchoni et al. (2019) use LASSO to target predictors in complete subset regressions (Elliott et al., 2013) (CSR) for forecasting consumer price inflation, stock market returns, industrial production growth, and employment growth. Elastic nets are used by Bork et al. (2020) to target predictors in partial least squares regressions for forecasting housing price growth, and by Borup and Schütte (2020) for targeting Google Trends series in predicting employment growth using RF, bagging, and CSR. LASSO, elastic nets, and Bayesian shrinkage have all been used for targeting factor models, e.g., for forecasting consumption and investment (Luciani, 2014) and gross domestic product growth and its subcomponents (Bulligan et al., 2015).

Given the demonstrated value of targeting, e.g., in regressions and factor-based analyses, it is natural to investigate the potential role of targeting of predictors in RF. This is important, as it addresses how to utilize RF in high dimensions. The challenge is that the performance of the RF algorithm historically has been considered extremely difficult to analyze (see, e.g., the discussion in Biau and Scornet, 2016). Most theoretical work on RF (Wager and Walther, 2016; Wager and Athey, 2018; Oprescu et al., 2019) requires the number of covariates to be small relative to sample size, and only few theoretical results are available. Biau (2012) demonstrates in a simplified setting how to achieve faster convergence than the usual $n^{-2/(p+2)}$ nonparametric regression rate in case of sparsity. Wager and Athey (2018) establish asymptotic normality for honest forests. Scornet et al. (2015) prove $L^2$ consistency of an RF algorithm close to Breiman's original specification.

Still, it remains unexplained how RF achieves its impressive predictive success in high-dimensional regimes, and this creates a gap between theory and practice. Whether RF benefits from a prior dimension reduction step in a high-dimensional setting is an important element of this discussion.

In this paper, we aim to reduce the gap by providing a theoretical and empirical assessment of the effects of targeting within the framework of RF. Our results are easy to grasp and highlight the components of the algorithm that are particularly impacted by targeting. We conduct our analysis in the following three steps.

First, we examine the ability of RF to detect a relatively small number of important predictors to split along when building trees in a high-dimensional setting with many potentially irrelevant predictors. Such a setting is motivated, e.g., by the empirical literature on asset return forecasting in which a plethora of predictors has been suggested (Welch and Goyal, 2008; Rapach and Zhou, 2013) or macroeconomic forecasting, in which, frequently, a large set of predictors is applied (Stock and Watson, 2002; McCracken and Ng, 2016).

We cast the analysis in terms of the probability $\rho$ of splitting along strong predictors. On the one hand, it is vital for the strength of individual trees that the important (strong) predictors are selected most of the time, i.e., $\rho$ must be sufficiently large. On the other hand, to control the variance of RF, it is important to randomize splitting directions when growing its trees, i.e., $\rho$ should not be so large as to jeopardize the benefits of averaging across trees in the forest. We establish lower and upper bounds on $\rho$ and show that the width of the interval for $\rho$ shrinks to zero as sample size increases. Based on this interval, we determine that the degree of feature sampling (the selection of splitting variables by RF), one of the few tuning parameters of RF, primarily controls the upper bound.

However, as established through simulations, $\rho$ will often be considerably below this bound in finite samples. Thus, feature sampling by itself does not ensure that $\rho$ is at an appropriate level, and hence the role for targeting. We show that the lower bound on $\rho$ is determined by two quantities. The first reflects the finite sample Classification And Regression Tree (CART) impurity decrease estimation error, and the second shows the maximal signal among the strong predictors. Through these quantities, we show that targeting can be used to lift the lower bound on $\rho$, primarily by reducing the CART estimation error. Thus, targeting can be used actively as a complement to feature sampling to secure an adequate probability of splitting along strong predictors.

Second, we show that the strength of individual decision trees in the forest is always improved by (proper) targeting, and we quantify this gain analytically in specific cases. In particular, for a linear regression function, we obtain explicit bounds on the mean squared error (MSE) of an ordinary tree, conditionally on the sequence of strong/weak splits, in terms of the MSEs of targeted trees with fewer leaves. From these conditional bounds and explicit expressions for the distribution of the underlying random variables, we derive bounds on the unconditional MSE of an ordinary tree. On this basis, we show that targeting can lead to significant gains in tree strength.

Third, we address the issue that, although tree strength is always improved by excluding weak predictors, the resulting TRF (targeted RF, i.e., RF with an initial targeting step) cannot be expected to perform uniformly better than ordinary RF, since the targeting step likely increases the correlation across individual trees in the forest. The inclusion of weak predictors can be seen as a way of injecting randomness into the tree-growing procedure, thereby increasing diversity across trees. More precisely, the expected number of so-called potential nearest neighbors increases with the dimension of the feature space, see Lin and Jeon (2006). Thus, targeting involves an inherent bias-variance trade-off—specifically, a tree strength-correlation trade-off, and the degree of targeting may be viewed as a tuning parameter.

We examine this strength-correlation trade-off in an extensive empirical analysis of the effects of targeting. An additional purpose of our empirical work is to assess the statistical and economic significance of the effect of targeting in typical areas of application. We consider two classical applications within different fields of economics and finance. The first is the challenging task of predicting the US equity premium in the setting of Welch and Goyal (2008). The second application considers the prediction of industrial production growth, employment growth, and consumer price inflation, using a large set of macroeconomic, financial, and sentiment variables from McCracken and Ng (2016). The former is chosen to illustrate an application where RF is particularly challenged theoretically due to a low signal-to-noise ratio, whereas the latter is chosen to highlight an application where RF is known to achieve desirable performance (Medeiros et al., 2019; Borup and Schütte, 2020). In line with Bai and Ng (2008), the set of targeted predictors is determined using LASSO regularization techniques. We choose LASSO as it is the most common regularization technique.

We synthesize our findings from the empirical analysis as follows. First, to address the strength-correlation trade-off inherent in targeting, we estimate the empirical MSE and correlation among trees. TRF performs well if a medium-sized subset of the initial predictors is targeted. In our applications, this amounts to targeting the best 10–30% of initial predictors. With too much targeting (too few predictors selected), the increased correlation between trees more than outweighs the gains in MSE from targeting. On the other hand, with too little targeting (too many predictors selected), the marginal decorrelation from including more predictors is more than outweighed by the loss in individual tree strength (increase in MSE).

Second, in terms of the significance of gains in our macroeconomic and financial applications, TRF performs particularly well for long forecast horizons. The prediction problem at long horizons is often challenging with limited signal (Galbraith and Tkacz, 2007), hence rendering TRF particularly useful in such cases. Third, TRF generates gains in predictive accuracy of substantial magnitude, up to 12–13%, relative to ordinary RF both in expansions and recessions.

The rest of the paper is laid out as follows. Section 2 starts with a mathematical introduction to both ordinary and targeted RF. This is followed by an analysis the ability of a forest to automatically detect good predictors and of the immediate gain in terms of tree strength from targeting. Section 3 presents empirical results on the effects of targeting in financial and macroeconomic applications. Section 4 concludes. All proofs are deferred to the Appendix. Throughout the paper we will be using the notation $\lfloor \cdot \rfloor$, $\lceil \cdot \rceil$, $\text{Leb}(\cdot)$, $\log_2(\cdot)$, and $\leq_{st}$ for the floor function, ceiling function, Lebesgue measure, logarithm to base 2, and first order stochastic dominance, respectively. Furthermore, we define the binomial coefficient $\binom{n}{k} := \frac{n!}{k!(n-k)!}$ for $n \geq k \geq 0$ and the Cartesian product $A \times B := \{(x, y) : x \in A, y \in B\}$ of any two sets $A$ and $B$.

## 2. The effect of targeting strong predictors in random forests

In this section, we concisely present ordinary and targeted RF. This is followed by an analysis of the ability of trees in the forest to select strong predictors over weak ones. This is a key property in high-dimensional settings. If the probability of splitting on strong variables is small, then it can be highly beneficial, or even necessary, to include an initial targeting step to avoid severe curse of dimensionality issues. In the subsequent analysis, we take a one-sided view on the effect of proper targeting and analyze the strength of an ordinary tree relative to a targeted one. While this does not yield a definitive performance comparison of RF versus TRF, it quantifies how much one would have to gain in terms of tree diversity to justify not targeting. In the empirical applications in Section 3, we assess this strength-diversity trade-off by estimating tree correlations and MSE for different levels of targeting.

### 2.1. The random forest algorithm and targeting

Given a sample of size $n$ of the form $\mathscr{D}_n = \{(X_1,Y_1),\ldots,(X_n,Y_n)\}$ with $(X_i,Y_i) \in \mathscr{X} \times \mathbb{R}$, $\mathscr{X} \subseteq \mathbb{R}^p$, $p$ indicating the number of predictors of $Y_i$ in the vector $X_i$, and $\mathbb{E}[Y^2] < \infty$, an RF produces a nonparametric estimate $\bar{f}_n = \bar{f}_n(\cdot;\mathscr{D}_n) \colon \mathscr{X} \to \mathbb{R}$ of the regression function $f := \mathbb{E}[Y|X = \cdot]$. It is an ensemble learning algorithm obtained by bagging, say, $B$ regression trees (the base learners) and can thus be represented as

$$\bar{f}_n(x) = \frac{1}{B}\sum_{b=1}^{B} \widehat{f}_n^b(x;\mathscr{D}_n), \qquad (1)$$

with $\widehat{f}_n^b(x;\mathscr{D}_n)$ the $b$th tree in the forest. Trees are assumed to be grown by the same set of rules, and their diversity is caused by injected (exogenous) randomness only. More precisely, $\widehat{f}_n^b(x;\mathscr{D}_n) = \widehat{f}_n(x;\mathscr{D}_n,\Theta_b)$, with $\Theta_1,\ldots,\Theta_B$ i.i.d. replicates of some random variable $\Theta$, which, e.g., can include decisions on resampling, splitting directions, and positions of splits. A tree $\widehat{f}_n(x;\mathscr{D}_n,\Theta)$ is a particular case of a partitioning estimate with feature space $\mathscr{X}$ partitioned into, say, $L$ hyper-rectangles (nodes), $(A_{i,n})_{i=1}^{L}$. The partition, which can depend on both $\mathscr{D}_n$ and $\Theta$, is constructed recursively by starting from $\mathscr{X}$ and performing a sequence of splits, each one perpendicular to the axes. Given a point $x \in \mathscr{X}$, the resulting tree estimate of $f(x)$ is the local average over the $Y_i$ for which the associated $X_i$ is in the same node as $x$, that is,

$$\widehat{f}_n(x;\mathscr{D}_n,\Theta) = \sum_{i=1}^{L} \bar{Y}_n(A_{i,n})\mathbb{1}_{A_{i,n}}(x), \qquad x \in \mathscr{X}, \qquad (2)$$

with $\bar{Y}_n(A) = \frac{1}{|\{k \colon X_k \in A\}|}\sum_{k \colon X_k \in A} Y_k$, and the convention $\bar{Y}_n(A) = 0$ if none of the observations belong to $A$. Many different specifications of $\Theta$ have been considered in the literature depending on whether the focus is on computational efficiency, adaptivity to high-dimensional feature spaces (many predictors), predictive strength, or analytic/theoretical tractability, see the discussion in Biau and Scornet (2016).

We use a typical tree-growing mechanism corresponding to a variant of Breiman's RF (Breiman, 2001). Each tree in the forest is based on a bootstrap sample from $\mathcal{D}_n$ with replacement. Splits are recursively performed in nodes until either (i) a maximal depth is reached, or (ii) splitting the node will imply that one of the child nodes contains strictly fewer bootstrap points than a certain threshold. We follow the conventional CART methodology (Breiman et al., 1984) and choose the optimal split $(i^\star, \tau^\star)$ in a given node $A$ by maximizing the impurity decrease in $A$,

$$L_n(i,\tau,A) = \frac{1}{n} \sum_{j:\, X_j \in A} (Y_j - \bar{Y}_n(A))^2 - \frac{1}{n} \sum_{j:\, X_j \in A \cap \{x: x^{(i)} \leq \tau\}} (Y_j - \bar{Y}_n(A \cap \{x : x^{(i)} \leq \tau\}))^2$$
$$- \frac{1}{n} \sum_{j:\, X_j \in A \cap \{x: x^{(i)} > \tau\}} (Y_j - \bar{Y}_n(A \cap \{x : x^{(i)} > \tau\}))^2,$$

over $i \in \mathcal{M}_{try}$ and $\tau \in A^{(i)} := \{x^{(i)} : x \in A\}$. Here, $x^{(i)}$ refers to the $i$th coordinate in $x$, and $\mathcal{M}_{try} = \mathcal{M}_{try}(A)$ is a random subset of $[p] := \{1,\ldots,p\}$, of fixed cardinality $m(p) := |\mathcal{M}_{try}|$, which determines the set of feasible split directions in $A$. The default is $m(p) = p$ in the **sklearn** library in Python, $m(p) = \lfloor \sqrt{p} \rfloor$ in the **ranger** package in R, and $m(p) = \lceil p/3 \rceil$ in both the **randomForest** package in R and the **TreeBagger** class in MATLAB.

### 2.1.1. Strong predictors and the targeted random forest

Assume throughout for simplicity that $\mathcal{X} = [0,1]^p$. Further, the key assumption throughout is that of a sparse setting, i.e., the regression function $f(\cdot)$ is of the form

$$f(x) = g(x_\mathcal{S}), \qquad x \in [0,1]^p, \tag{3}$$

for some measurable function $g: [0,1]^s \to \mathbb{R}$ and subset $\mathcal{S} \subseteq [p]$, with $x_\mathcal{S} = (x^{(i)})_{i \in \mathcal{S}}$ and $s := |\mathcal{S}|$ significantly smaller than $p$. In order for $\mathcal{S}$ to be unique, we will assume that it is chosen as small as possible among those satisfying (3). Many applications have $p = 100$ or larger, and $s \approx 5$ or less, see, e.g., the discussions in Rapach and Zhou (2013) and Chinco et al. (2019). The predictors in $\mathcal{S}$ are referred to as strong, and the remaining as weak, in line with Breiman (2004), Biau (2012), and Biau and Scornet (2016). Other definitions of strong and weak predictors exist in the literature, e.g., in Ishwaran (2015), where a predictor is strong if it affects the conditional distribution of $Y$ given $X = x$ (in particular, any of its moments), and so the number of strong predictors will generally be larger in that setting. Syrganis and Zampetakis (2020) employ a strong sparsity assumption in their analysis of RF.

The targeted RF (henceforth TRF) algorithm is identical to the RF above, except that an initial step is included with the aim to filter out some of the weak predictors. The search is for a (relatively small) subset $\mathscr{S}' \subseteq [p]$ satisfying $\mathscr{S}' \supseteq \mathscr{S}$. The resulting targeted estimator of $f(\cdot)$ is thus constant along directions in $[p] \setminus \mathscr{S}'$, and its trees correspond to partitions of $[0,1]^{s'}$ with $s' := |\mathscr{S}'|$. This makes for an assumption of proper targeting, i.e., strong predictors are not discarded, which we need for some of our analysis. There are various ways of choosing the targeted set $\mathscr{S}'$. A typical choice is via regularization, such as LASSO or elastic net. In settings with very large $p$ relative to $s$, screening techniques may also be appealing, see, e.g., Fan and Lv (2008) and Fan et al. (2011). Fan and Lv (2010) review variable selection techniques in high dimensions, including both regularization and screening. Bayesian shrinkage may be applied, too. In the following theoretical exposition, we take as given a fixed but arbitrary targeting procedure and focus on its effects on RF.

*2.1.2. Assumptions*

We will impose one or more of the following assumptions:

(A1) The data $\mathscr{D}_n = \{(X_1, Y_1), \ldots, (X_n, Y_n)\}$ form an ergodic sequence.

(A2) The input vector $X$ is uniformly distributed on $\mathscr{X} = [0,1]^p$.

(A3) The regression function is linear, $f(x) = \beta_0 + \sum_{i \in \mathscr{S}} \beta_i x^{(i)}$.

Assumption (A1) is imposed to ensure that empirical averages converge to their theoretical counterparts. This assumption is mild and holds for most stationary processes. The second assumption, (A2), is classical in the nonparametric regression literature, and it is often sufficient to assume that the copula density of $X$ is bounded from above and below (Biau, 2012; Györfi et al., 2002; Scornet et al., 2015; Wager and Athey, 2018). Assumption (A3) is mainly imposed for the sake of simplicity and smooth exposition. Under this assumption, $\mathscr{S} = \{i : \beta_i \neq 0\}$. We emphasize that most of what follows could as well be worked out under the assumption that $f(\cdot)$ is additive, $f(x) = \sum_{i \in \mathscr{S}} f_i(x^{(i)})$. The additive regression framework is convenient when trees are built up on splits based on the CART criterion. Given a node $A$ of a tree, there will asymptotically (as $n \to \infty$) always be a split in $A$ leading to a decrease in impurity, unless $f(\cdot)$ is constant on $A$ (Scornet et al., 2015, Technical Lemma 1, Supplement). In contrast, if $f(\cdot)$ is not additive, there may be situations in which $f(\cdot)$ is non-constant on $A$, but the population counterpart of the CART criterion (given in (4)) is identically zero, even when $A = [0,1]^p$ and (A2) is satisfied. Consequently, in such a situation, it is very difficult to assess split positions (and, hence, the structure of trees) theoretically without imposing other strict assumptions. In sum, nonlinearities can be allowed for, yet to carry out a clear and smooth exposition we proceed under the above assumptions without much loss of generality.

## 2.2. *The probability of splitting on strong predictors*

The ability of a given algorithm to navigate in high dimensions depends on its probability of splitting on strong predictors. We analyze this splitting probability both with and without an initial targeting step. To cover both situations at once, consider selecting a general subset $\mathscr{A} \subseteq [p]$ of cardinality $a := |\mathscr{A}|$ prior to building the trees of the forest. The case $\mathscr{A} = [p]$ returns the ordinary RF, and $\mathscr{S} \subseteq \mathscr{A} \subsetneq [p]$ a proper TRF. Fix a node $A \subseteq [0,1]^a$ in a given tree, select at random $m(a) \leq a$ feasible split directions from the random subset $\mathscr{M}_{try}^{\mathscr{A}}$ of $\mathscr{A}$, and let $\rho_n(\mathscr{A})$ be the probability that a split in $A$ is performed along a strong predictor. Let $s(a) := |\mathscr{A} \cap \mathscr{S}|$ be the number of strong predictors in $\mathscr{A}$. If $s(a) \ll a$, then $m(a)$ may be tuned sufficiently low to ensure $\rho_n(\mathscr{A}) < 1$ and thereby induce tree diversity, resulting in a variance reduction of the RF estimator.

On the other hand, to avoid severely biased trees, it is important that splitting directions are not simply chosen at random, and that strong predictors are selected most of the time, which means that $\rho_n(\mathscr{A}) \gg s(a)/a$. This implies that $m(a)$ should not be tuned too low. This section aims to show that (i) it might not be possible to ensure that $\rho_n(\mathscr{A})$ is sufficiently large simply by tuning $m(a)$ high (close to $a$), and (ii) using (proper) targeting instead can increase $\rho_n(\mathscr{A})$.

While $\rho_n(\mathscr{A})$ is difficult to assess directly, we are able to provide useful bounds on this probability. Before stating the formal result, some general notation is introduced. Let $L^\star$ denote the population counterpart of the CART objective function (3), that is,

$$L^\star(i, \tau, A) = \text{Var}_A(Y) - \mathbb{P}_A(X^{(i)} \leq \tau)\text{Var}_A(Y \mid X^{(i)} \leq \tau) \\ - \mathbb{P}_A(X^{(i)} > \tau)\text{Var}_A(Y \mid X^{(i)} > \tau), \quad (4)$$

subscript $A$ indicating conditioning on $\{X \in A\}$. Define

$$\delta_n(\mathscr{A}) = \sup_{i \in \mathscr{A}, \tau \in A^{(i)}} |L_n(i,\tau) - L^\star(i,\tau)|, \quad \text{and} \quad C^\star(\mathscr{A}) = \sup_{i \in \mathscr{M}_{try}^{\mathscr{A}} \cap \mathscr{S}, \tau \in A^{(i)}} L^\star(i,\tau),$$

with the convention $\sup \emptyset = 0$ and suppressing the dependence on $A$ in $L_n$ and $L^\star$. The quantity $\delta_n(\mathscr{A})$ reflects the finite sample disturbances from the estimation of impurity decrease, and $C^\star(\mathscr{A})$ the maximal signal (relative to the CART criterion) among the strong predictors in $\mathcal{M}_{try}^{\mathscr{A}}$.

**Theorem 1.** *The probability $\rho_n(\mathscr{A})$ of splitting on a strong variable satisfies*

$$\mathbb{P}(2\delta_n(\mathscr{A}) < C^\star(\mathscr{A})) \leq \rho_n(\mathscr{A}) \leq \mathbb{P}(\mathcal{M}_{try}^{\mathscr{A}} \cap \mathscr{S} \neq \emptyset). \tag{5}$$

*Under (A1) and $\mathbb{E}[|Y|^\gamma] < \infty$ for some $\gamma > 2$, the impurity decrease estimation error is asymptotically negligible, $\delta_n(\mathscr{A}) \to 0$ as $n \to \infty$ with probability one. If, in addition, (A2)–(A3) are satisfied, then*

$$\rho_n(\mathscr{A}) \longrightarrow \mathbb{P}(\mathcal{M}_{try}^{\mathscr{A}} \cap \mathscr{S} \neq \emptyset), \qquad n \to \infty. \tag{6}$$

In general, the less the finite sample error (the smaller $\delta_n$) or the stronger the signal (the higher $C^\star$), the tighter is the interval (5) for $\rho_n(\mathscr{A})$, the probability of splitting on a strong variable. An additional mild moment condition on $Y$ ensures that the splitting probability approaches its upper bound. In fact, from the proof of the theorem, it follows that Assumptions (A2)–(A3) are not strictly needed for (6). Since $\delta_n(\mathscr{A}) \to 0$ with probability one under (A1) and the moment condition, the only additional property needed is that

$$\sup_{\tau \in A^{(i)}} L^\star(\tau, i) > 0, \tag{7}$$

for each $i \in \mathscr{S}$. This is satisfied, e.g., if $f(\cdot)$ is additive, $f(x) = \sum_{i \in \mathscr{S}} f_i(x^{(i)})$, with $f_i(\cdot)$ continuous and not constant on $A^{(i)}$ (Scornet et al., 2015, Technical Lemma 1, Supplement), thus relaxing (A3). Similarly, for additive $f(\cdot)$, it is possible to relax (A2) by using the c.d.f. to transform the predictors (Davis and Nielsen, 2020, Lemma 1). The transformed predictors are not necessarily independent, but as noted in Section 2.1.2, it is often sufficient that the copula density is bounded from above and below. We will not go into further details with this, but instead focus on the possibility of controlling the splitting probability through the bounds from Theorem 1.

### 2.2.1. Control of upper bound

The estimators (both RF and TRF) can be tuned to ensure diversity of trees by lowering the upper bound in (5) in Theorem 1 through the choice of the function $m(\cdot)$, thereby forcing $\rho_n(\mathscr{A})$ away from one. To elaborate on this, the upper bound can be calculated explicitly, using the hypergeometric distribution, as

$$\mathbb{P}\big(\mathcal{M}_{try}^{\mathcal{A}} \cap \mathcal{S} \neq \emptyset\big) = 1 - \mathbb{1}_{\{m(a) < a - s(a)\}} \binom{a - s(a)}{m(a)} \bigg/ \binom{a}{m(a)}. \tag{8}$$

As long as $a$ is large compared to $s(a)$, proper choice of $m(\cdot)$ ensures that $\rho_n(\mathcal{A})$ is not too large. To give an example, a representative forecasting exercise with $a = 40$ predictors, of which $s(a) = 5$ strong, and splitting direction chosen from among $m(a) = \lceil a/3 \rceil = 14$ feasible directions would produce an upper bound of $\mathbb{P}(\mathcal{M}_{try}^{\mathcal{A}} \cap \mathcal{S} \neq \emptyset) = 0.9$, low enough to ensure randomness in the splitting procedure while still much larger than $s(a)/a = 0.125$. In particular, if $\rho_n(\mathcal{A})$ is close to its upper bound which occurs asymptotically by (6) in Theorem 1, the ability of the trees to select strong predictors to split on is, in principle, controllable through choice of $m(\cdot)$. However, as we show in simulations below, $\rho_n(\mathcal{A})$ can be far below the upper bound in small samples, i.e., the interval (5) can be wide. Thus, additional tools, besides the choice of $m(\cdot)$, are needed to ensure that $\rho_n(\mathcal{A})$ is not too small and hence the role for targeting.

### 2.2.2. Control of lower bound

Targeting of predictors can be used to raise the lower bound in (5) in Theorem 1. First, given $\mathcal{M}_{try}^{\mathcal{A}}$, the quantity $C^\star(\mathcal{A})$ is deterministic and may for some classes of $f(\cdot)$ be computed explicitly. For instance, under (A3) it can be verified that

$$C^\star(\mathcal{A}) = \frac{1}{16} \sup_{i \in \mathcal{M}_{try}^{\mathcal{A}} \cap \mathcal{S}} \beta_i^2 \, \text{Leb}(A^{(i)})^2, \tag{9}$$

with $A = A^{(1)} \times \cdots \times A^{(p)}$ being the node to split and $A^{(i)}$ corresponding to the feasible interval for the $i$th predictor (see also Biau, 2012, Section 3). In general, it holds that $C^\star(\mathcal{A}) \leq \sup_{i \in \mathcal{A}, \tau \in A^{(i)}} L^\star(i, \tau)$. Here, the right-hand side is independent of $m(a)$ and can, for some regression functions $f(\cdot)$, be arbitrarily small, even for fixed $A$ and $\text{Var}(f(X))$ (see Proposition 2 below). This means that the dependence of the lower bound $\mathbb{P}(2\delta_n(\mathcal{A}) < C^\star(\mathcal{A}))$ on $m(a)$ is limited. While the presence of $\delta_n(\mathcal{A})$ in the bound is unavoidable, its magnitude can be reduced simply by considering a smaller set of predictors, $\mathcal{B}$. If, in addition, $C^\star$ is roughly the same for $\mathcal{A}$ and $\mathcal{B}$, the lower bound increases. Ultimately, this means that $\rho_n(\mathcal{B})$ is forced to exceed a larger bound than $\rho_n(\mathcal{A})$. The following proposition gives sufficient conditions for one targeting set $\mathcal{B}$ to be preferred over another set $\mathcal{A}$.

**Proposition 1.** *Let $\mathcal{A}, \mathcal{B} \subseteq [p]$ such that $\delta_n(\mathcal{B}) \leq_{st} \delta_n(\mathcal{A})$ and $C^\star(\mathcal{A}) \leq_{st} C^\star(\mathcal{B})$. Then the lower bound on $\rho_n$ is larger for $\mathcal{B}$ than for $\mathcal{A}$, that is,*

$$\mathbb{P}(2\delta_n(\mathscr{A}) < C^\star(\mathscr{A})) \leq \mathbb{P}(2\delta_n(\mathscr{B}) < C^\star(\mathscr{B})). \tag{10}$$

*In particular, using a targeting set $\mathscr{A}$ increases the lower bound on the probability $\rho_n$ of splitting on strong predictors if $C^\star([p]) \leq_{st} C^\star(\mathscr{A})$.*

Intuitively, the condition $\delta_n(\mathscr{B}) \leq_{st} \delta_n(\mathscr{A})$ means that $\mathscr{B} \subseteq \mathscr{A}$. Under suitable assumptions, extreme value theory implies an approximate relation

$$\mathbb{P}(\delta_n(\mathscr{A}) \geq x) \approx a\mathbb{P}(Z_n \geq x), \tag{11}$$

for $s(a) \ll a \ll n$ and some random variable $Z_n$ not depending on $\mathscr{A}$. This suggests that $\delta_n(\mathscr{B}) \leq_{st} \delta_n(\mathscr{A})$ as long as the cardinality of $\mathscr{B}$ is smaller than that of $\mathscr{A}$ (both being much larger than $s$). The condition $C^\star(\mathscr{A}) \leq_{st} C^\star(\mathscr{B})$ is related to the likelihood of having strong predictors in $\mathscr{M}_{try}^{\mathscr{A}}$ relative to $\mathscr{M}_{try}^{\mathscr{B}}$. It is of no use simply to discard arbitrary predictors.

Consider the case that all directions are feasible ($m(\cdot)$ is the identity function), and $\mathscr{B}$ contains the same strong directions as $\mathscr{A}$. This implies that $C^\star(\mathscr{A})$ and $C^\star(\mathscr{B})$ are deterministic and $C^\star(\mathscr{A}) = C^\star(\mathscr{B}) = C^\star$. Thus, if all directions are feasible, and $\mathscr{A}$ represents a targeted set that does not eliminate strong predictors from the non-targeted set of all original predictors, $[p]$, then targeting increases the lower bound on $\rho_n$. If the approximation (11) applies, a rough estimate of the gain from targeting (increase in the lower bound in Theorem 1) is

$$\mathbb{P}(2\delta_n(\mathscr{A}) < C^\star) - \mathbb{P}(2\delta_n([p]) < C^\star) \approx (p-a)\mathbb{P}(2Z_n \geq C^\star). \tag{12}$$

Thus, targeting of predictors in RF improves the lower bound on the probability of splitting on strong predictors roughly linearly in the number of weak predictors discarded. This indicates that the ability of TRF to navigate in high-dimensional settings can dominate that of ordinary RF.

### 2.2.3. Sampling experiments

By Theorem 1, two key objects, the finite sample impurity decrease estimation error $\delta_n(\mathscr{A})$ and the maximal signal $C^\star(\mathscr{A})$, determine the lower bound on the probability $\rho_n(\mathscr{A})$ of splitting on strong variables. By Proposition 1, using targeting to push these two quantities in a stochastic order sense can increase the lower bound. It is of interest to know whether the impact carries over to $\rho_n(\mathscr{A})$ directly in finite samples. Thus, we design simulation studies to explicitly assess the partial effect of $\delta_n = \delta_n([p])$ and $C^\star = C^\star([p])$ on $\rho_n = \rho_n([p])$. Throughout the simulations, we set $m(p) = p$, $\mathscr{S} = \{1\}$, and let $Y$ given $X = x$ be Gaussian with variance $\sigma^2 > 0$. In particular, $C^\star$ is deterministic, and $\rho_n \to 1$ as $n \to \infty$, by Theorem 1. For simplicity, we take $A = [0,1]^p$ and normalize $f(\cdot)$ so that $\text{Var}(f(X)) = 1$. The signal-to-noise ratio (SNR) may be expressed simply in terms of $\sigma^2$,

$$\text{SNR} := \frac{\text{Var}(f(X))}{\text{Var}(Y)} = \frac{1}{1 + \sigma^2}. \tag{13}$$

Under Assumptions (A2)–(A3), the restrictions used in the simulations imply $|\beta_1| = \sqrt{12}$ and $\beta_2 = \cdots = \beta_p = 0$, since $\mathscr{S} = \{1\}$ and the variance of a uniform on $[0,1]$ is $1/12$, i.e., $\text{Var}(f(X)) = 1 = \beta_1^2/12$.

**The effect of $\delta_n$ on $\rho_n$**  We consider variations in the SNR (13) and the number of weak predictors as these are key drivers of $\delta_n$. The higher the SNR or the lower the number of weak predictors, the lower the value of $\delta_n$. We focus on the linear case (A3), $\beta_1 = \sqrt{12}$. Given $n$, $p$, and SNR, we repeat the following experiment 10,000 times to obtain a Monte Carlo estimate of $\rho_n$:

- Generate $n$ realizations $(X_1, Y_1), \ldots, (X_n, Y_n)$.

- Compute the value $L_n(i, X_j^{(i)})$ of the CART objective function (3) at $\tau = X_j^{(i)}$, for $i = 1, \ldots, p$ and $j = 1, \ldots, n$.

- Assign the value 1 to the current repetition of the experiment if $L_n(i, X_j^{(i)})$ is largest for $i = 1$, and 0 otherwise.

Figure 1 shows the resulting (approximate) probability $\rho_n$ as a function of SNR for different values of $n$ and $p$. Consistently with our theoretical results, $\rho_n$ increases as $p$ is reduced or SNR or $n$ is increased. A challenging predictive environment, with low SNR or many weak predictors, reduces the probability of splitting on strong predictors considerably in finite samples. This relates directly, e.g., to financial applications aiming at predicting stock, bond, or exchange rate returns. Here, $n = 100$ would be a typical sample size, and likely SNR $\leq 0.1$, since asset returns contain a sizeable amount of inherently unpredictable variation (Rapach and Zhou, 2013). Moreover, a plethora of predictors exists. If, say, around five percent of the predictors are strong, the sampling experiment suggests that the probability of splitting on strong predictors is only of the order one half.

**The effect of $C^\star$ on $\rho_n$**  Given the normalization $\text{Var}(f(X)) = 1$ and linearity (A3), it follows from (9) that $C^\star = 3/4$. However, for nonlinear $f(\cdot)$, this value can be very different.

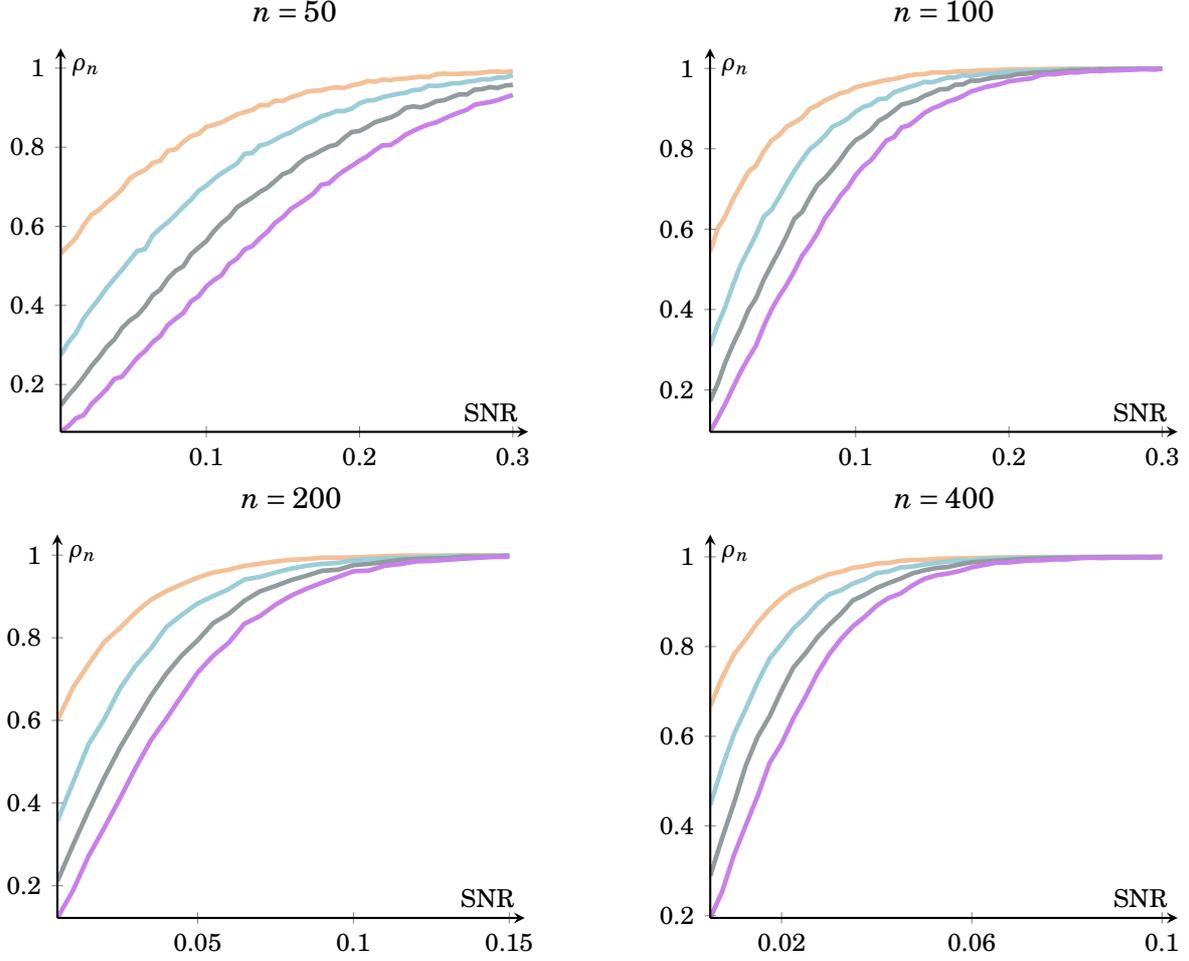

**Figure 1:** Probability of splitting on a strong predictor as a function of SNR (I)
This figure shows the probability of splitting on the strong predictor $X^{(1)}$ as function of SNR for different sample sizes $n$ and number of predictors $p = 2$ (orange), $p = 4$ (blue), $p = 8$ (gray), and $p = 16$ (purple).

If $f(x)$ tends to fluctuate around a certain level for $x$ in some region $A$, the gain from placing a split in $A$ can be limited. To illustrate, consider an oscillating regression function

$$f(x) = \sqrt{2}\sin(\alpha x^{(1)}), \qquad x = (x^{(1)}, \ldots, x^{(p)})' \in [0,1]^p, \tag{14}$$

with frequency $\alpha$ a multiple of $2\pi$ for simplicity. This relates to applications involving seasonality with $X^{(1)}$ indicating calendar time and predicting, e.g., industrial production or retail sales (Ghysels and Osborn, 2001). Given the amplitude of $\sqrt{2}$ in the example, $f(\cdot)$ does indeed satisfy $\text{Var}(f(X)) = 1$ under Assumption (A2). Further, $\alpha$ controls the maximal signal, $C^\star$, as shown in the following proposition.

**Proposition 2.** *For $f$ given by* (14) *and under Assumption (A2), the maximal signal satisfies* $C^\star \leq 4(\alpha - 2)^{-1}$.

Thus, the higher the frequency $\alpha$, the less is the gain from placing a split along the

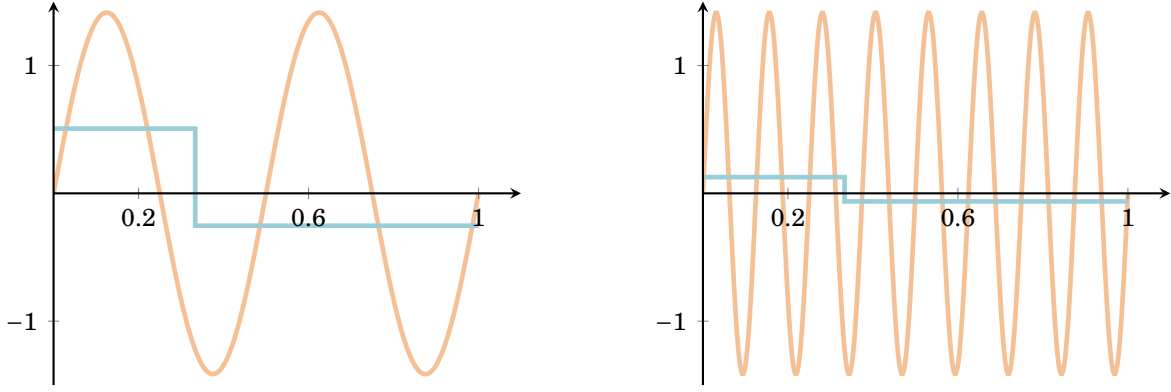

**Figure 2:** The best approximating step function

This figure shows the best approximating step function $x \mapsto a\mathbb{1}_{[0,\tau]}(x) + b\mathbb{1}_{(\tau,1]}(x)$ to (14) in case $\alpha = 4\pi$ (left) and $\alpha = 16\pi$ (right), for $\tau = 1/3$.

strong variable, in terms of impurity decrease. Even for moderate values of $\alpha$, the signal is much smaller than the value 3/4 for the linear specification. For instance, if $\alpha = 4\pi$, then $C^\star \leq 0.3786$, i.e., the signal is at most about half of the linear one. Consequently, we should expect $\rho_n$ to be considerably smaller in this example. With oscillations on both sides of a split position $\tau$, the best approximating step function $x \mapsto a\mathbb{1}_{[0,\tau]}(x) + b\mathbb{1}_{(\tau,1]}(x)$ of $f$ has $a, b \approx 0$, so the improvement over the zero function is limited. This is illustrated in Figure 2. While there will be few oscillations in, say, $[0, \tau]$ for $\tau$ close to 0, such a split leads to a modest decrease in impurity, since $X^{(1)}$ only falls into $[0, \tau]$ for a small fraction of observations. The question is whether the effect on $C^\star$ carries over to $\rho_n$. Based on Proposition 2 and this discussion, we expect $\rho_n$ to decrease in $\alpha$ and, generally, be much smaller than the values in Figure 1 for the linear specification of $f(\cdot)$. We design a new simulation experiment along the lines of the preceding one, restricting attention to the case $p = 8$. We compare the linear regression function to (14) for various values of $\alpha$ and $n$.

Results of the experiment are presented in Figure 3. Generally, the oscillating behavior of the regression function harms the ability of the CART criterion to detect the signal, and the more oscillations, the worse the performance. In particular, when $\alpha = 16\pi$, $\rho_n$ remains below 0.2 for all values of SNR and $n$ considered, which is not much better than choosing the split direction at random ($s/p = 0.125$).

While the lack of identification of strong predictors is particularly prevalent for oscillating regression functions, the issue remains for less chaotic specifications. To illustrate, consider $f(x) \propto (x^{(1)} - 1/2)^2$ and the piecewise polynomial, which is due to Györfi et al. (2002),

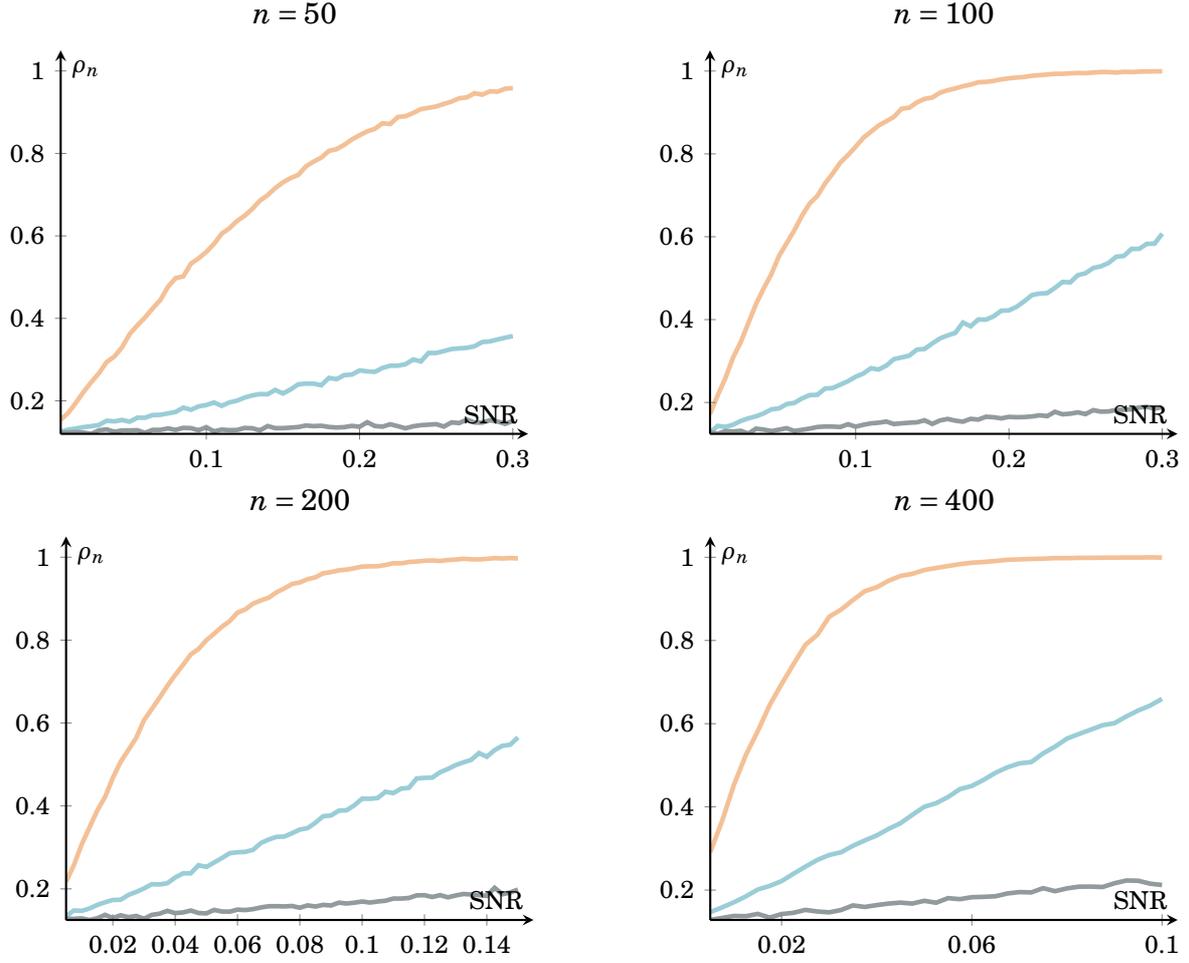

**Figure 3:** Probability of splitting on a strong predictor as a function of SNR (II)
This figure shows the probability of splitting on the strong predictor $X^{(1)}$ as function of SNR in the linear case (orange) compared to the oscillating (14), with $\alpha = 4\pi$ (blue) and $\alpha = 16\pi$ (gray). In all cases, $p = 8$.

$$f(x) \propto \begin{cases} (2x^{(1)} + 1)^2/2 & \text{if } x^{(1)} \in [0, 1/4), \\ x^{(1)} + 3/8 & \text{if } x^{(1)} \in [1/4, 1/2), \\ -5(2x^{(1)} - 6/5)^2 + 43/40 & \text{if } x^{(1)} \in [1/2, 3/4), \\ 2x^{(1)} - 7/8 & \text{if } x^{(1)} \in [3/4, 1]. \end{cases} \quad (15)$$

The value of $C^\star$ is computed numerically to 0.3125 for the second-order polynomial and 0.2565 for (15), both considerably smaller than 0.75, the value for $f(\cdot)$ linear. Although these cases are less extreme than the oscillating, results from a third sampling experiment, shown in Figure 4, demonstrate that the probability of selecting the strong predictor $X^{(1)}$ is reduced significantly relative to the linear case.

Taken together, the results show that increased sample size, fewer weak predictors,

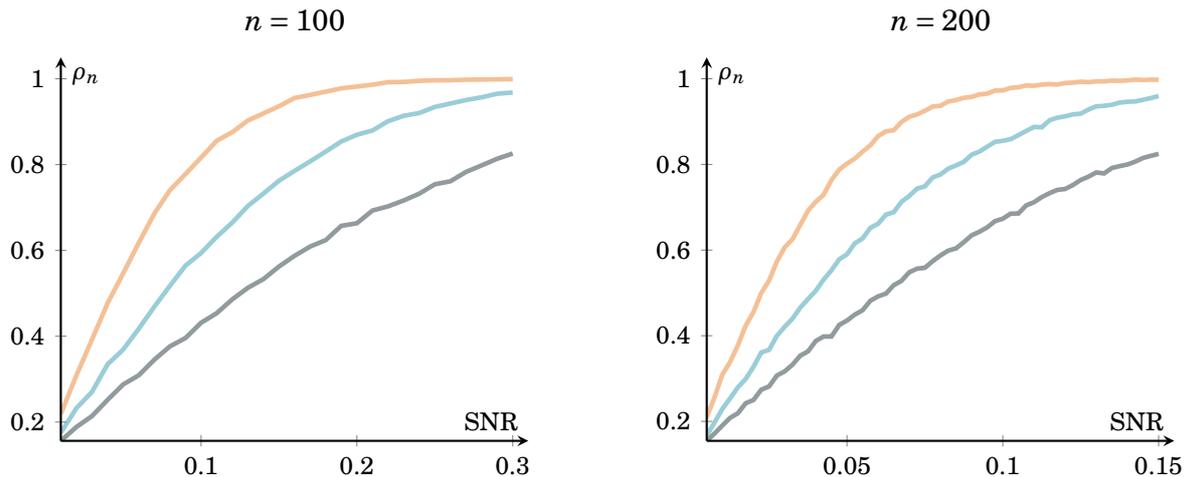

**Figure 4:** Probability of splitting on a strong predictor as a function of SNR (III)
This figure shows the probability of splitting on the strong predictor $X^{(1)}$ as function of SNR in the linear setting (orange) compared to the cases $f(x) \propto (x^{(1)} - 1/2)^2$ (blue) and $f(\cdot)$ given by the piecewise polynomial (15) (gray). In all cases, $p = 8$.

and stronger signal-to-noise ratio, or, more generally, reduced finite sample error in the estimation of impurity decrease and increased maximal signal can increase both the lower bound on the probability of splitting on strong variables and the probability itself in finite samples.

*2.3. Strength of trees*

In the following analysis, we consider the effect of targeting in terms of strength of individual shallow trees; that is, trees whose "size" does not increase with sample size $n$. To evaluate the strength of a forest, one would need to take into account both the strength and the diversity of its trees. In this section, we investigate the gain from targeting in terms of tree strength, thus showing how much must be gained from tree diversification to justify not targeting. For simplicity of exposition, we consider a particular type of shallow trees, in which the number of leaves is fixed, and nodes are split in a best-first fashion. Such trees are widely applied in practice and implemented in most programming languages. For example, in Python's **sklearn** library, these trees are tuned by the *max_leaf_nodes* parameter.

In our empirical applications in Section 3, we use trees of fixed depth—that is, trees for which an exact number of edges must be traversed to reach any of their leaves. With only a few modifications to the splitting rule, we conjecture that the following analysis can be carried out for trees of fixed depth as well.

Recall that, given data $\mathscr{D}_n$, an ordinary tree forecast $\widehat{f}_{L,n}(x)$ of $f(x)$ with $L$ leaves takes

the form

$$\widehat{f}_{L,n}(x) = \sum_{i=1}^{L} \bar{Y}_n(A_{i,n})\mathbb{1}_{A_{i,n}}(x), \qquad x \in [0,1]^p, \qquad (16)$$

with $(A_{i,n})_{i=1}^{L}$ a partition of $[0,1]^p$, which in Breiman's (Breiman, 2001) algorithm depends on $\mathscr{D}_n$ and the random input selection $\Theta$ (assuming no bootstrap step). For simplicity, assume $\mathscr{S} = \{1\}$ and perfect targeting, $\mathscr{S}' = \mathscr{S}$. Thus, the targeted tree forecast $\widetilde{f}_{L,n}(x)$ is of the same form (16), but with splits only along the strong predictor. Hence, $(A_{i,n})_{i=1}^{L}$ is replaced by a partition $(B_{i,n})_{i=1}^{L}$ of $[0,1]$ and corresponding local averages,

$$\widetilde{f}_{L,n}(x) = \sum_{i=1}^{L} \bar{Y}_n(B_{i,n})\mathbb{1}_{B_{i,n}}(x^{(1)}), \qquad x \in [0,1]^p. \qquad (17)$$

We assess the effect of targeting by comparing the strength of $\widetilde{f}_{L,n}$ and $\widehat{f}_{L,n}$. Given $L \ll n$, we assume that the partition $(A_{i,n})_{i=1}^{L}$ can be built in a theoretically optimal way, that is, $(A_{i,n})_{i=1}^{L} = (A_i^L)_{i=1}^{L}$ is obtained by starting from $A_1^1 = [0,1]^p$ and then applying the following recipe recursively:

- For $k < L$, let $\mathscr{M}_{try}^{(k)}$ be a randomly chosen subset of $[p]$ of size $m$.

- Pick a node $A$ in $(A_i^k)_{i=1}^{k}$. If $\mathscr{M}_{try}^{(k)} \cap \mathscr{S} \neq \emptyset$, the $k$th split $(i^\star, \tau^\star)$ is determined by optimizing (4) over $i \in \mathscr{M}_{try}^{(k)} \cap \mathscr{S}$ and $\tau \in A^{(i)}$.

- The partition $(A_i^{k+1})_{i=1}^{k+1}$ is the same as $(A_i^k)_{i=1}^{k}$ except from $A$, which is divided into $A \cap \{x : x^{(i^\star)} \leq \tau^\star\}$ and $A \cap \{x : x^{(i^\star)} > \tau^\star\}$, with $(i^\star, \tau^\star)$ the chosen split.

In particular, the probability $\rho$ of splitting along a strong variable coincides with the upper bound from (5) in Theorem 1, explicitly given as (see (8))

$$\rho = 1 - \mathbb{1}_{\{m+s<p\}} \binom{p-s}{m} \Big/ \binom{p}{m}. \qquad (18)$$

In case the optimum $(i^\star, \tau^\star)$ is not unique, we assume that a certain deterministic tie-breaking rule is employed. The construction outlined above is similar to the one used in practice, except that we assume that $L^\star(\cdot)$ can be optimized, rather than $L_n(\cdot)$. The node $A$ to split in a given step is determined in a best-first fashion:

(R) The $k$th split is performed in the node leading to maximal impurity decrease, i.e., if $(i_j, \tau_j)$ is the optimal split in $A_j^k$, then the node to split is $A = A_{j^\star}^k$, with $j^\star = \mathrm{argmax}_j L^\star(i_j, \tau_j, A_j^k)$.

The partition $(B_i)_{i=1}^L$ is obtained similarly, but all splits are placed in $\mathscr{S}$. Define the corresponding partition-optimal trees $\widehat{f}_L$ and $\widetilde{f}_L$ by

$$\widehat{f}_L(x) = \sum_{i=1}^L \mathbb{E}[Y \mid X \in A_i]\mathbb{1}_{A_i}(x), \quad \text{and} \quad \widetilde{f}_L(x) = \sum_{i=1}^L \mathbb{E}[Y \mid X^{(1)} \in B_i]\mathbb{1}_{B_i}(x^{(1)}). \qquad (19)$$

Under wide conditions, $\widehat{f}_{L,n}$ and $\widetilde{f}_{L,n}$ from (16) and (17) converge to their partition-optimal counterparts (see, e.g., Wager and Walther, 2016; Davis and Nielsen, 2020), meaning that their relative performance can be assessed through (19) for large $n$. Thus, we restrict attention to $\widehat{f}_L$ and $\widetilde{f}_L$ in the following.

As an example of the structure of $\widehat{f}_L$ and $\widetilde{f}_L$, consider $p = 2$ and $L = 6$, and let $Z_k = \mathbb{1}_{\{\mathcal{M}_{try}^{(k)} \cap \mathscr{S} \neq \emptyset\}}$ indicate whether the $k$th split is performed along the strong predictor or not. Consider the case that $Z_k = 0$ for $k \in \{1, 4\}$ and $Z_k = 1$ otherwise, and denote by $\tau_1, \tau_2, \tau_3 \in [0, 1]$ and $\gamma_1, \gamma_2 \in [0, 1]$ the corresponding splits along the strong and weak predictor, respectively. This implies that the first and fourth splits are along the weak predictor for RF. The first split $\gamma_1$ leads to a partition of $\widehat{f}_2$ given by $A_1^2 = [0,1] \times [0, \gamma_1]$ and $A_2^2 = [0,1] \times (\gamma_1, 1]$. The next split will be at $\tau_1$, in either $A_1^2$ or $A_2^2$, depending on the ranking of ties, since both result in an impurity decrease of $\beta_1^2/16$ under assumption (A3) (see (9)). Moreover, the optimal split is always at the midpoint, so $\tau_1 = 1/2$. The next strong split will for sure be placed in the other node, at $\tau_2 = 1/2$. The fourth split is weak and will be placed somewhere in one of the four possible nodes, the specific position being determined by the tie-breaking rule. Finally, each of the last five nodes leads to the same maximal impurity decrease, so we will again rely on the ranking of ties. As before, $\tau_3$ is placed at the midpoint which, depending on the node, is either 1/4 or 3/4.

For the targeted tree, the first split is at $\tau_1^* = 1/2$, then $\tau_2^* \in \{1/4, 3/4\}$, and $\tau_3^* = \{1/4, 3/4\} \setminus \{\tau_2^*\}$. The fourth and fifth splits are placed at the midpoints in the resulting intervals of the strong predictor. This means that the ordinary tree gets to split only at two distinct places along the strong predictor (at $\tau_1$ and $\tau_3$), although three strong splits are placed, while the targeted tree splits five times along the strong predictor. An example of the two partitions is given in Figure 5, where the horizontal axis represents a strong predictor and the vertical axis represents a weak predictor. The fourth and fifth splits are represented by a dashed line to indicate the additional splits on the strong predictor when the number of leaves is fixed at $L = 6$.

The example shows that the order in which strong and weak splits are placed can have

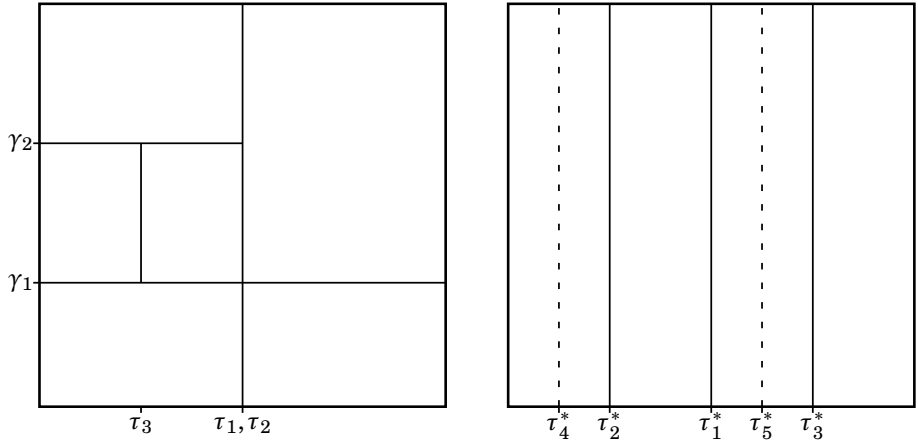

**Figure 5:** Ordinary and targeted partitions

This figure shows a partition of $[0,1]^2$ for an ordinary tree of size $L = 6$ (left) and its targeted counterpart (right). The horizontal and vertical axes correspond to a strong and a weak predictor, respectively. Only two distinct splits on the strong predictor are performed in the ordinary tree.

a significant impact on tree strength. This is formalized more generally in the next theorem which provides bounds on the strength of the ordinary tree $\widehat{f}_{L,n}$ relative to the corresponding targeted tree $\widetilde{f}_{L,n}$. Write $\mathrm{MSE}(\widehat{f}) := \mathbb{E}[(f(X) - \widehat{f}(X))^2]$ for an estimator $\widehat{f} = \widehat{f}(\,\cdot\,; \mathscr{D}_n)$ of $f$ which is independent of $X$, let the notation $\mathrm{MSE}_\Theta$ indicate that probabilities are computed conditionally on $\Theta$, and define the functions

$$\underline{\iota}(x) = 2^{\lfloor \log_2(x) \rfloor} \qquad \text{and} \qquad \bar{\iota}(x) = 2^{\lceil \log_2(x) \rceil} \tag{20}$$

which satisfy $\underline{\iota}(x) \leq x \leq \bar{\iota}(x)$ with equalities when $x = 2^k$ for some integer $k$.

**Theorem 2.** *Suppose that Assumption (A2)–(A3) are satisfied, and that the splitting rule (R) is employed. Then*

$$\mathrm{MSE}(\widetilde{f}_L) = \beta_1^2 \frac{7\underline{\iota}(L) - 3L}{48\underline{\iota}(L)^3}, \qquad L \geq 1, \tag{21}$$

*and*

$$\mathrm{MSE}_\Theta\left(\widetilde{f}_{\underline{\iota}\left(\ell_0 + \frac{N - \ell_0 + 1}{\ell_0(L - N)}\right)}\right) \geq \mathrm{MSE}_\Theta(\widehat{f}_L) \geq \mathrm{MSE}_\Theta\left(\widetilde{f}_{\bar{\iota}\left(1 + \frac{N}{\ell_1}\right)}\right), \qquad L \geq 1, \tag{22}$$

*with* $\Theta = (\mathscr{M}_{try}^{(1)}, \ldots, \mathscr{M}_{try}^{(L-1)})$, $N = \sum_{k=1}^{L-1} \mathbb{1}_{\{\mathscr{M}_{try}^{(k)} \cap \mathscr{S} \neq \emptyset\}}$, *and* $\ell_i = \min\{k : \mathbb{1}_{\{\mathscr{M}_{try}^{(k)} \cap \mathscr{S} \neq \emptyset\}} = i\}$.

In the event that the set $\{k : \mathbb{1}_{\{\mathscr{M}_{try}^{(k)} \cap \mathscr{S} \neq \emptyset\}} = i\}$ considered in Theorem 2 is empty, we set $\ell_i = 1$. Thus, $N$ is the number of strong splits among $L - 1$ possible, while $\ell_0$ and $\ell_1$ refer to the first time a weak split and a strong split are placed, respectively. In the example above, and in Figure 5, $N = 5$, $\ell_0 = 1$, and $\ell_1 = 2$. The following heuristic arguments provide intuition for the bounds (22) in the theorem.

**Lower bound** Before placing any strong splits, the tree is partitioned into $\ell_1$ subtrees, without improving MSE. The splitting rule (R) implies that the $N$ strong splits are roughly equally distributed across subtrees, meaning that none of these can be better than a targeted tree with $1 + N/\ell_1 \leq L$ leaves, using that $\ell_1 \geq 1$ and $N \leq L-1$.

**Upper bound** The first $\ell_0 - 1$ splits are strong, so up to this point, the ordinary tree $\widehat{f}_L$ is identical to the targeted tree with $\ell_0$ leaves. However, in $\widehat{f}_L$, these $\ell_0$ nodes are expanded into subtrees. The worst possible subtree would be one which is first divided into $L - N$ branches by weak splits among them and then receives a number of strong splits. Again, the splitting rule (R) ensures that each branch in this subtree receives roughly $\frac{N-\ell_0+1}{\ell_0(L-N)}$ strong splits. Since this number is on top of the initial $\ell_0 - 1$ strong splits, $\widehat{f}_L$ is no worse than a targeted tree with $\ell_0 + \frac{N-\ell_0+1}{\ell_0(L-N)} \leq L$ leaves, using that $N - \ell_0 + 1 \geq 0$, $\ell_0(L-N) \geq 1$, and $N \leq L - 1$.

The heuristic arguments, using only rough knowledge of the distribution of the strong splits, are in Theorem 2 turned into the rigorous bounds in (22), using the functions $\underline{\iota}$ and $\bar{\iota}$ from (20). From the proof of Theorem 2 it follows that the bounds (22) can be obtained for a much more general specification of the regression function $f(\cdot)$ than the linear one imposed by (A3). Specifically, by modifying the splitting rule (R) slightly and always placing splits at the midpoint of the interval $A^{(1)}$ (the projection of the node to be split, $A$, onto the strong direction) when $\mathcal{M}_{try}^{(k)} \cap \mathcal{S} \neq \emptyset$ rather than optimizing (4), it is sufficient to assume that $f(x) = g(x^{(1)})$ for a general continuous function $g(\cdot)$ which is non-constant on any open interval of $[0,1]$. In other words, nonlinearities can be allowed in this setting.

While Theorem 2 provides bounds conditionally on $\Theta$, these can be translated into explicit bounds on the unconditional MSE of the ordinary tree. Let $g_0(x,y) = \mathrm{MSE}\bigl(\widetilde{f}_{\underline{\iota}(y + \frac{x-y+1}{y(L-x)})}\bigr)$ and $g_1(x,y) = \mathrm{MSE}\bigl(\widetilde{f}_{\bar{\iota}(1+\frac{x}{y})}\bigr)$ be the functions determining the bounds (22) for given $N$, $\ell_0$, and $\ell_1$. These functions can be evaluated using (21) from Theorem 2. The following corollary provides tractable expressions for the distributions of $(N, \ell_0)$ and $(N, \ell_1)$ and explicitly bounds the MSE of the ordinary tree above and below by averages over targeted trees with fewer leaves.

**Corollary 1.** *Under the conditions of Theorem 2,*

$$\mathbb{E}[g_0(N,\ell_0)] \geq \mathrm{MSE}(\widehat{f}_L) \geq \mathbb{E}[g_1(N,\ell_1)], \qquad (23)$$

*in which the probability mass functions of $(N, \ell_0)$ and $(N, \ell_1)$ are given by*

$$\mathbb{P}(N=n, \ell_0 = k)$$

$$= \begin{cases} \rho^{k-1}(1-\rho)\operatorname{Bin}(n+1-k; L-1-k, \rho) & \textit{if } n \in (0, L-1) \textit{ and } k \in [1, n+1] \\ \rho^n(1-\rho)^{L-1-n} & \textit{if } n \in \{0, L-1\} \textit{ and } k = 1 \\ 0 & \textit{otherwise,} \end{cases}$$

*respectively*

$$\mathbb{P}(N=n, \ell_1 = k)$$

$$= \begin{cases} \rho(1-\rho)^{k-1}\operatorname{Bin}(n-1; L-1-k, \rho) & \textit{if } n \in (0, L-1) \textit{ and } k \in [1, L-n] \\ \rho^n(1-\rho)^{L-1-n} & \textit{if } n \in \{0, L-1\} \textit{ and } k = 1 \\ 0 & \textit{otherwise,} \end{cases}$$

*with $\rho$ defined in (18) and $\operatorname{Bin}(k; n, \rho) = \binom{n}{k}\rho^k(1-\rho)^{n-k}$ the probability mass function of the Binomial distribution with n trials and success probability $\rho$.*

From (23) and $1 + N/\ell_1 \leq L$, it follows that $\operatorname{MSE}(\widehat{f}_L) > \operatorname{MSE}(\widetilde{f}_{i(L)})$, i.e., there is an immediate gain in tree strength from targeting (again, $\operatorname{MSE}(\widehat{f}_L) > \operatorname{MSE}(\widetilde{f}_L)$ for $L = 2^k$). Figure 6 illustrates the bounds on $\operatorname{MSE}(\widehat{f}_L)$ from (23) in the corollary by showing $\mathbb{E}[g_0(N, \ell_0)]$ and $\mathbb{E}[g_1(N, \ell_1)]$ as functions of $\rho$ for two different values of $L$. The probability mass functions provided by the corollary are used to compute these functions. In the figure, these best and worse case scenarios for the performance of the ordinary tree are compared to that of the targeted tree, $\operatorname{MSE}(\widetilde{f}_L)$. With only one strong predictor, $\rho$ will not exceed 0.5 when the total number of predictors is $p \geq 2$ and the cardinality of $\mathcal{M}_{try}$ is set to the default value, $m = \lceil p/3 \rceil$. For the values of $L$ considered here, the MSE for the ordinary tree, located somewhere in the shaded region in Figure 6, is much larger than for its targeted counterpart, shown in the bottom part of the figure, even for large values of $\rho$.

## 3. Empirical applications

In this section, we investigate the effect of targeting predictors in random forests empirically by assessing the relative predictive ability of ordinary versus targeted RF in two classical applications involving many initial predictors, namely, equity premium prediction in the style of Welch and Goyal (2008), and the prediction of industrial production growth, employ-

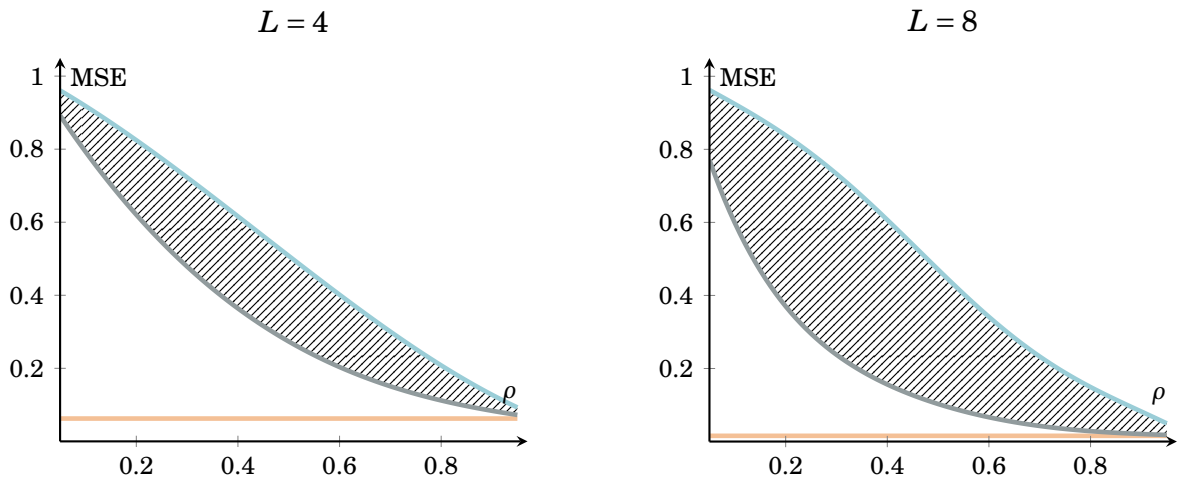

**Figure 6:** Comparison of MSE to upper and lower bounds

This figure shows a comparison of MSE($\widetilde{f}_L$) (orange) to the upper (blue) and lower (gray) bounds of MSE($\widehat{f}_L$) from Corollary 1 as functions of $\rho$, with $\beta_1 = \sqrt{12}$, for two values of $L$. The graph of MSE($\widehat{f}_L$) is located somewhere in the shaded region.

ment growth, and consumer price inflation from a large panel of monthly macroeconomic, financial, and sentiment variables as in Stock and Watson (2002), using the McCracken and Ng (2016) data. The first application illustrates a case in which RF is particularly challenged theoretically due to low SNR, and the second a case in which RF is known to achieve superior performance (Medeiros et al., 2019).

We consider standard transformations of the data for which ergodicity (A1) can reasonably be assumed. We expect that most of our theoretical results apply more generally than under the stated assumptions of uniformity (A2) and linearity (A3). We focus first on the linear case in the applications and then turn to nonlinearities in Section 3.3.1.

Following standard practice in the literature, we evaluate the accuracy of our point forecasts based on squared prediction errors, and we use the Diebold-Mariano or DM (Diebold and Mariano, 1995) test to compare the performance of TRF against ordinary RF. The null hypothesis is that the forecast from TRF does not outperform that from ordinary RF, and the (one-sided) alternative is that it does. The DM test statistic is constructed using HAC standard errors and a Bartlett kernel with lags truncated at $h-1$, with $h$ being the forecast horizon. A maximum tree depth of 3 is applied, e.g., corresponding to the median of values used in Gu et al. (2020). The implementation is done in Python using the standard **sklearn** library. For the $b$th tree in (1), $\Theta_b$ specifies the bootstrap draw of size $n$ from $\mathscr{D}_n$ with replacement and for each node the random selection $\mathscr{M}_{try}^{\mathscr{A}}$ of size $m(a) = \lceil a/3 \rceil$ of feasible split directions from $\mathscr{A}$, with $a = |\mathscr{A}|$, i.e., $\mathscr{A} = [p]$ and $a = p$ for ordinary RF, $a < p$ for TRF.

## 3.1. Applications with many initial predictors

The financial and macroeconomic applications are described in Sections 3.1.1 and 3.1.2, respectively. Section 3.2 outlines our implementation of targeting, and Section 3.3 presents empirical results based on monthly data for predictive ability one month ahead ($h = 1$), one quarter ahead ($h = 3$), and one year ahead ($h = 12$).

### 3.1.1. Predicting the equity premium

Our first application, the prediction of the monthly equity premium in the spirit of Welch and Goyal (2008), is based on a long tradition in finance. The forecasting objective is the return to the US stock market in excess of the risk-free rate, i.e., the equity premium. For the stock market index $P_t$, the logarithmic return is defined by $R_{t+h} = \log P_{t+h} - \log P_t$ and the equity premium by

$$Y_t = R_{t+h} - R^f_{t+h}, \qquad (24)$$

with $R^f_{t+h}$ being the continuously compounded risk-free rate of return. We use the S&P500 month-end cum dividend index returns from 1960–2017 for $R_{t+h}$ and the monthly Treasury bill rate for $R^f_{t+h}$. Following Welch and Goyal (2008), we aim to predict $Y_t$ using the most prominent predictors $X_t$ in the literature. We include the dividend-price ratio, dividend-earnings ratio, earnings-price ratio, dividend-yield ratio, book-to-market ratio, net equity expansion, Treasury bill rate, term spread, default return spread, default yield spread, long-term rate of return, long-term yield, stock variance, and inflation for a total of $p = 14$ predictors. Their construction follows Welch and Goyal (2008), and data on $X_t$ and $Y_t$ are obtained from Amit Goyal's website.[*]

As in Welch and Goyal (2008), we consider an expanding window estimation scheme, including all data up to the point in time at which the forecast is constructed. The initial estimation window spans the period from 1960 through 1974, so that the first forecast is generated for the $h$th month of 1975 and the last for December 2017.

---

[*] http://www.hec.unil.ch/agoyal/.

*3.1.2. Predicting macroeconomic variables*

Our macroeconomic application covers monthly industrial production growth, employment growth, and consumer price inflation prediction. We treat the industrial production index (*IP*) and employment (*EMP*) as $I(1)$ series, and the consumer price index (*CPI*) as an $I(2)$ series, following Stock and Watson (2002) and McCracken and Ng (2016). The forecasting object is the logarithmic difference or cumulative growth for a given horizon $h$,

$$Y_t = \log Z_{t+h} - \log Z_t, \tag{25}$$

for $Z_t = \{IP_t, EMP_t\}$. For our third macroeconomic forecasting object, we consider the second difference of the logarithm of $CPI_t$, i.e., $CPI$ acceleration, and accumulate this over the $h$-step horizon.

Following Stock and Watson (2002), we aim to predict $Y_t$ using a large panel of monthly predictors. We use the McCracken and Ng (2016) data, which contains a broad set of macroeconomic, financial, and sentiment variables, including data for constructing $Y_t$.[†] We restrict the sample period to 1970-2018, as most series become available from 1970, and remove those with missing values during the period. This yields a set of $p = 100$ predictors. We transform the data as proposed by McCracken and Ng (2016).

As in the financial application, we consider an expanding window estimation scheme. The initial estimation window runs from 1970 through 1984, so the first forecast is for the $h$th month of 1985.

## 3.2. Targeting predictors

There are various ways of choosing the targeted set $\mathcal{S}'$. In this application, we follow Bai and Ng (2008) and consider the LASSO estimator $\widehat{\beta}^\lambda$ of the linear regression coefficients, obtained as

$$(\widehat{\alpha}^\lambda, \widehat{\beta}^\lambda) = \underset{\alpha,\beta}{\operatorname{argmin}} \sum_{i=1}^{n}(Y_i - \alpha - \beta' X_i)^2 + \lambda \|\beta\|_{\ell_1}, \tag{26}$$

with $\|\cdot\|_{\ell_1}$ the $\ell_1$ norm. The minimization problem (26) corresponds to the Lagrangian for the minimization of the sum of squared errors over a rhomboid $\{\beta : \|\beta\|_{\ell_1} \leq C\}$, which in turn is the reason that $\widehat{\beta}^\lambda$ will often be sparse, i.e., it will only have few non-zero entries. The choice of $\lambda$ or, equivalently, $C$, controls the degree of sparsity. Under suitable conditions, $\lambda = \lambda(n)$ can be tuned in such a way that the LASSO asymptotically identifies the true sparsity pattern $\mathscr{S}$ as $n \to \infty$ (Hastie et al., 2015, Ch. 11). Thus, we choose the targeted set $\mathscr{S}' = \mathscr{S}'(\lambda)$ as

$$\mathscr{S}' = \{i \in [p] : \widehat{\beta}_i^\lambda \neq 0\}, \tag{27}$$

and use $\lambda$ to control the number of selected predictors, $s'$. An important feature of $\mathscr{S}'$ is that the selection is explicitly based on the predictors' ability to forecast $Y$. Effectively, the degree of targeting is a tuning parameter of the procedure. For example, Bai and Ng (2008) target 30 predictors in their factor-based setting. In the RF context, the degree of targeting is important for the performance of the resulting TRF, as we show in the following.

---

[†]https://research.stlouisfed.org/econ/mccracken/fred-databases/.

### 3.3. Results

Table 1 reports the ratio of the mean squared prediction error (MSE) of TRF for various numbers of targeted predictors, $s'$, to that of ordinary RF, both for the full out-of-sample period and, to gauge the economic significance, for NBER dated recessions and expansions with the forecasting objective in either of the two states. Reported values below unity indicate superior performance of TRF relative to RF and are highlighted with bold. Significance levels of the DM test are displayed using standard notation with three, two, and one asterisks representing $p$-values less than 1%, 5%, and 10%, respectively.

Results are reported for a broad range of $s'$, the number of predictors targeted in TRF. In the macroeconomic applications, $\lambda$ in (27) is tuned to target $s' = 5, 10, 20, 30,$ or $50$ predictors in TRF against the $p = 100$ available to RF. These values span a very sparse setting with only a few (five) targeted predictors, medium-dimensional settings in which several predictors are left out, yet 10–30 are targeted, and one with only half of the predictors discarded. In the financial application, $\lambda$ is tuned to $s' = 2, 5,$ or $10$ predictors in TRF compared to $p = 14$ for RF, similarly representing sparse and medium settings as well as a case with only a small share of initial predictors discarded. From the theory, targeting always increases the strength of individual trees (Section 2.3) but might come at a cost of reduced diversity across trees, hence lowering the benefit of the variance reduction inherent in RF. Here, we analyze this trade-off empirically along with the magnitude of the impact of targeting. In the following, we synthesize our findings.

**Targeted RF performs particularly well for long horizons** At the longest forecasting horizon considered, $h = 12$, TRF delivers improvements over ordinary RF over the full

**Table 1:** Predictive ability of RF versus TRF

| | Full out-of-sample | | | NBER recessions | | | NBER expansions | | |
|---|---|---|---|---|---|---|---|---|---|
| $s'$ | $h=1$ | $h=3$ | $h=12$ | $h=1$ | $h=3$ | $h=12$ | $h=1$ | $h=3$ | $h=12$ |
| | *Panel A: Employment growth* | | | | | | | | |
| 5 | 1.031 | **0.993** | **0.909**** | **0.975** | **0.983** | **0.920*** | 1.035 | **0.995** | **0.906*** |
| 10 | **0.994** | **0.996** | **0.878**** | **0.927** | **0.868**** | **0.906*** | **0.999** | **0.994** | **0.870**** |
| 20 | **0.992** | **0.996** | **0.920**** | 1.018 | **0.908**** | **0.987** | **0.990** | 1.004 | **0.901**** |
| 30 | **0.996** | **0.994** | **0.916**** | 1.015 | **0.968*** | **0.994** | **0.994** | **0.997** | **0.894**** |
| 50 | 1.025 | **0.996** | **0.926**** | 1.094 | **0.969**** | 1.019 | 1.019 | **0.998** | **0.900**** |
| | *Panel B: Industrial production growth* | | | | | | | | |
| 5 | 1.095 | 1.076 | **0.872** | 1.319 | 1.204 | 1.032 | 1.010 | 1.009 | **0.751*** |
| 10 | 1.009 | 1.055 | **0.927** | 1.134 | 1.188 | 1.028 | **0.962** | **0.985** | **0.851*** |
| 20 | 1.021 | 1.001 | **0.875**** | 1.124 | 1.066 | **0.946** | **0.979** | **0.966** | **0.821*** |
| 30 | 1.012 | **0.984** | **0.898**** | 1.067 | 1.077 | **0.954*** | **0.991** | **0.935*** | **0.856*** |
| 50 | 1.013 | **0.960** | **0.889*** | 1.022 | 1.099 | **0.918**** | 1.010 | **0.886**** | 0.867 |
| | *Panel C: Consumer price inflation (acceleration)* | | | | | | | | |
| 5 | **0.993** | **0.906** | **0.986** | **0.925** | **0.822** | **0.890*** | 1.022 | **0.974** | 1.013 |
| 10 | **0.986** | **0.922** | **0.984** | **0.989** | **0.845** | **0.924**** | **0.986** | **0.983** | 1.010 |
| 20 | 1.006 | **0.915** | **0.993** | 1.006 | **0.830** | **0.924**** | 1.006 | **0.983** | 1.012 |
| 30 | 1.005 | **0.926*** | **0.988** | 1.016 | **0.867*** | **0.944*** | 1.004 | **0.974** | 1.000 |
| 50 | 1.010 | **0.964*** | **0.988** | 1.011 | **0.931*** | **0.976**** | 1.010 | **0.990** | **0.991** |
| | *Panel D: Equity premium (S&P500 Index returns)* | | | | | | | | |
| 2 | 1.011 | 1.099 | 1.056 | **0.961** | 1.212 | 1.097 | 1.005 | 1.094 | 1.047 |
| 5 | **0.979** | 1.015 | **0.941*** | **0.913**** | 1.084 | 1.123 | **0.988** | 1.000 | **0.922*** |
| 10 | 1.005 | **0.991** | **0.955**** | **0.993** | 1.004 | 1.004 | 1.010 | **0.978*** | **0.943**** |

This table reports the ratio of mean squared prediction error (MSE) of each version of TRF to that of ordinary RF, both for the full out-of-sample period and for the NBER dated recessions and expansions with the forecasting objective belonging to either of the two states. Across the macroeconomic (employment growth, industrial production growth, and consumer price inflation) and financial applications, forecast horizons $h = 1, 3, 12$ are considered. The penalty $\lambda$ is tuned to target $s'$ predictors. Bold indicates values of relative MSE below unity and thus improvements from targeting. Superscripts ***, **, and * indicate statistical significance, based on a (one-sided) Diebold-Mariano test statistic using HAC standard errors with a Bartlett kernel of bandwidth $h-1$, at significance levels 1%, 5%, and 10%, respectively.

out-of-sample period. MSE ratios are below unity both for the equity premium and the three macroeconomic variables as well as for all degrees of targeting considered, except in a single case—the equity premium, heavy targeting. Improvements from targeting are statistically significant at level 10% or better in most cases. The same general pattern applies for $h = 12$ during recessions and expansions, the only exception during expansions being inflation, and during recessions—the equity premium. The results are interesting, as the forecasting of macroeconomic variables is known to be particularly challenging at long horizons. For example, Galbraith and Tkacz (2007) find deteriorating predictability for increasing forecasting horizons across a large set of macroeconomic and financial variables, and Galbraith (2003) documents a similar pattern for GDP and inflation, specifically. Thus, TRF delivers significant improvements in settings in which predictability is generally challenging. This corresponds well with the results from Section 2.2.3, showing that a low SNR can reduce the probability of splitting on strong variables considerably, and that targeting of predictors reduces the curse of dimensionality issues involved. It is likely that SNR is particularly low for long-horizon macroeconomic forecasting, which therefore explains the significant gains to targeting in these cases.

**RF can yield substantial gains in predictive accuracy**  For employment growth, $CPI$, and the equity premium, and for each of the three forecasting horizons over the full out-of-sample period, TRF improves over ordinary RF in most cases with at most two-three exceptions across the different numbers of targeted predictors considered. The gain in predictive accuracy is considerable in many cases, up to 12–13%, which is noteworthy, considering that the benchmark model, RF, is a sophisticated forecasting technique itself. Whenever TRF is not preferred over RF, the difference is typically small and economically negligible. For instance, for consumer price inflation, this amounts to 0.5–1% over the full period. For industrial production growth, most gains occur at the longest horizon, $h = 12$, again with magnitudes up to 12–13% but with no improvements at $h = 1$, hence reinforcing the particular relevance of targeting in low SNR environments.

**Targeted RF yields improvements in both expansions and recessions**  For employment growth, gains in predictive accuracy of TRF over RF occur both during NBER dated recessions and expansions. They appear strongest during recessions at the quarterly forecasting horizon and during expansions at the yearly. For industrial production growth, considerable gains to targeting are observed during expansions, while gains are relatively more frequent during recessions for consumer price inflation. The improvements in predictive accuracy for the equity premium occur both during recessions and expansions. Like for employment growth, they are strongest at long horizons during expansions, while during recessions they are strongest at shorter horizons (monthly, in case of the equity premium, quarterly for employment growth).

**A medium degree of targeting is preferred**  For employment growth, over the full period, the only cases in which TRF produces less predictive accuracy than ordinary RF are for very light ($s' = 50$) or very heavy ($s' = 5$) targeting at the short horizon ($h = 1$). A similar pattern is seen for the equity premium with TRF not beating RF for $s' = 10$ or 2 at the short horizon. During expansions, an intermediate degree of targeting is preferred at the short horizon for employment and industrial production growth. The notion that an optimal degree of targeting exists is meaningful. With too much screening, the improvements in predictive performance for a given tree are insufficient to offset the large increase in correlation among trees, and the variance reduction from averaging the forest never fully kicks in.

In terms of the analysis in Section 2.2, although the probability of splitting on strong predictors is improved, the low number of predictors targeted renders trees very similar. On the other hand, with too little screening targeting does not sufficiently increase the probability of splitting on strong predictors as too many week predictors are retained.

The empirical results are consistent with the analysis in Section 2, i.e., improvements in tree strength come at the cost of increased correlation across trees, thus limiting the reduction in variance from averaging across trees. Moreover, the targeting step is naturally at risk of discarding strong variables. To analyze this inherent trade-off in targeting empirically, we follow Breiman (2001) and estimate the MSE and (pairwise) correlation among trees in the forest. Specifically, conditionally on two randomization parameters $\Theta$ and $\Theta'$, define the covariance $\kappa(\Theta,\Theta')$ between the (prediction) errors of the corresponding trees,

$$\kappa(\Theta,\Theta') = \mathbb{E}\big[\big(Y - \widehat{f}_n(X;\mathscr{D}_n,\Theta)\big)\big(Y - \widehat{f}_n(X;\mathscr{D}_n,\Theta')\big) \mid \Theta,\Theta',\mathscr{D}_n\big]. \tag{28}$$

The MSE of a tree built from $\mathscr{D}_n$, averaged over $\Theta$, is

$$\text{MSE}(\text{tree}) = \mathbb{E}\big[\kappa(\Theta,\Theta) \mid \mathscr{D}_n\big], \tag{29}$$

and the correlation $\overline{\rho}$ between tree errors is

$$\overline{\rho} = \frac{\mathbb{E}\big[\kappa(\Theta,\Theta') \mid \mathscr{D}_n\big]}{\mathbb{E}\big[\sqrt{\kappa(\Theta,\Theta)} \mid \mathscr{D}_n\big]^2}, \tag{30}$$

with $\Theta$ and $\Theta'$ independent. We estimate the quantities in (29) and (30) by applying out-of-sample analogs in the following way. Using the same initial in-sample period of 15 years as in Table 1, we estimate 500 individual trees, i.e., trees that would compose an ordinary RF. For each tree, we predict the outcome of interest for the entire out-of-sample period. To estimate the expectation over $\Theta$, we average over trees, and to estimate the expectation over data, we average across the out-of-sample predictions. We repeat this procedure for the full range of targeting levels $s' = 1,\dots,p$, with $s' = p$ corresponding to ordinary RF.

Figure 7 presents the estimated quantities in (29) and (30), i.e., average out-of-sample MSE and correlation among trees grown, for each version of TRF, including ordinary RF. We focus here on the case of $h = 1$ month ahead prediction of employment growth noted previously, i.e., $p = 100$. From the figure, targeting evidently involves a trade-off between tree MSE and tree correlation. As the number of targeted predictors is increased, MSE generally increases, consistently with the analysis in Section 2.3. At the same time, tree correlation decreases (right panel of figure). Indeed, for very few targeted predictors, tree correlation increases rapidly, almost exponentially. Similarly, MSE decreases fast for few targeted predictors. The results in Table 1 suggest, however, that the increase in correlation is too strong to justify the decrease in MSE from heavy targeting, i.e., TRF predictions are inferior to ordinary RF predictions for extreme degrees of targeting. Similar trade-offs apply for industrial production growth, inflation, and the equity premium.

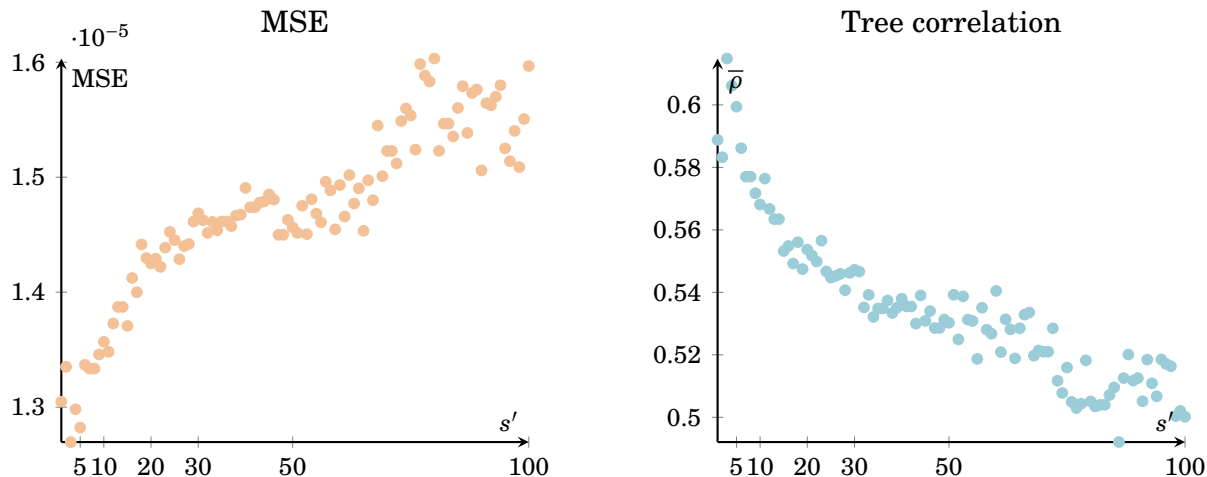

**Figure 7:** Tree strength-correlation trade-off
This figure shows the MSE (left figure) and tree correlation (right figure) for TRF over the range $s' = 1,\ldots,100$, with $s' = p = 100$ corresponding to the ordinary RF. Values shown are for the application to one month ahead employment growth prediction.

It stands out that an intermediate degree of targeting balances the trade-off between tree strength and correlation optimally. Across our applications, TRF mostly delivers improvements for a medium-sized set of targeted predictors amounting to the best 10–30% of the commonly applied set of initial predictors.

### 3.3.1. Targeting and nonlinearities

Next, we investigate whether our conclusions regarding the effect of targeting predictors on the predictive accuracy of RF carry over to the case of an extended targeting step, accommodating that predictors may be strong only when included nonlinearly, e.g., as squared terms or interactions. When targeting predictors in their original form via LASSO, there is a risk of throwing away relevant information that RF would find useful. Thus, for the macroeconomic applications, we now conduct the targeting on an expanded set of

**Table 2:** Predictive ability of RF versus TRF with expanded feature set in targeting

| $s'$ | Full out-of-sample | | | NBER recessions | | | NBER expansions | | |
|---|---|---|---|---|---|---|---|---|---|
| | $h=1$ | $h=3$ | $h=12$ | $h=1$ | $h=3$ | $h=12$ | $h=1$ | $h=3$ | $h=12$ |
| | *Panel A: Employment growth* | | | | | | | | |
| 5  | 1.034    | 1.021    | **0.917**\*\* | **0.905** | 1.011       | **0.927**    | 1.043       | 1.022       | **0.914**\*   |
| 10 | 1.006    | **0.991** | **0.876**\*\* | 1.030    | **0.919**\*\* | **0.910**\*\* | 1.005       | **0.997**   | **0.866**\*\* |
| 20 | **0.993** | **0.992** | **0.901**\*\* | 1.002    | **0.949**\*\* | **0.983**    | **0.992**   | **0.996**   | **0.878**\*\* |
| 30 | **0.987**\*\* | **0.990** | **0.921**\*\* | **0.994** | **0.914**\*\*\* | 1.018       | **0.987**\*\* | **0.997**   | **0.894**\*\* |
| 50 | **0.994** | **0.994** | **0.926**\*\* | **0.994** | **0.941**\*\*\* | 1.019       | **0.994**   | **0.999**   | **0.901**\*\* |
| | *Panel B: Industrial production growth* | | | | | | | | |
| 5  | 1.095    | 1.076    | **0.868**    | 1.319    | 1.204       | 1.029        | 1.010       | 1.008       | **0.747**\*   |
| 10 | 1.013    | 1.056    | **0.932**    | 1.152    | 1.190       | 1.031        | **0.960**   | **0.986**   | **0.857**\*   |
| 20 | **0.996** | **0.999** | **0.884**\*\* | 1.044    | 1.060       | **0.985**    | **0.978**   | **0.967**   | **0.821**\*\* |
| 30 | **0.997** | 1.007    | **0.931**    | 1.075    | 1.090       | 1.056        | **0.967**   | **0.964**   | **0.808**\*   |
| 50 | **0.999** | 1.005    | **0.908**\*  | 1.079    | 1.181       | 1.014        | **0.968**\* | **0.912**\*\* | **0.837**\*   |
| | *Panel C: Consumer price inflation (acceleration)* | | | | | | | | |
| 5  | 1.023    | **0.906** | **0.992**    | 1.015    | **0.827**   | **0.889**\*  | 1.027       | **0.969**   | 1.020        |
| 10 | **0.980** | **0.923**\* | 1.019        | **0.938** | **0.878**   | 1.019        | **0.998**   | **0.959**\*\* | 1.019        |
| 20 | **0.989** | **0.934**\* | 1.011        | **0.980** | **0.891**   | 1.018        | **0.993**   | **0.969**\*\* | 1.010        |
| 30 | **0.985** | **0.934**\* | **1.006**    | **0.960** | **0.884**   | 1.036        | **0.996**   | **0.975**\* | **0.997**    |
| 50 | **0.999** | **0.958**\* | **0.995**    | 1.003    | **0.927**\* | **0.969**\*\*\* | **0.996**   | **0.983**   | 1.002        |
| | *Panel D: Equity premium (S&P500 Index returns)* | | | | | | | | |
| 2  | 1.038    | **0.969** | 1.021        | 1.135    | **0.948**   | **0.733**\*  | 1.048       | 1.019       | 1.006        |
| 5  | **0.981** | 1.011    | **0.939**\*  | **0.903**\*\* | 1.074  | 1.141        | **0.989**   | 1.004       | **0.918**\*   |
| 10 | 1.004    | **0.993** | **0.949**\*\* | **0.989** | 1.007      | 1.003        | 1.007       | **0.981**\* | **0.937**\*\* |

This table reports the ratio of mean squared prediction error (MSE) of each version of TRF to that of ordinary RF, both for the full out-of-sample period and for the NBER dated recessions and expansions with the forecasting objective in either of the two states. Across the macroeconomic (employment growth, industrial production growth, and consumer price inflation) and financial applications, forecast horizons $h = 1, 3, 12$ are considered. The penalty $\lambda$ is tuned to target $s'$ predictors. Bold indicates values of relative MSE below unity and thus improvements from targeting. Superscripts \*\*\*, \*\*, and \* indicate statistical significance based on a (one-sided) Diebold-Mariano test statistic using HAC standard errors with a Bartlett kernel of bandwidth $h-1$, at significance levels 1%, 5%, and 10%, respectively.

predictors, comprised of the original predictors and their squared and cubed terms. We still feed the original predictors to the RF, as this is supposed to identify the relevant nonlinear transforms itself and to keep all else unaltered.

For TRF, after targeting, we feed the targeted predictors in their original form to the algorithm for the final prediction step, even if they were selected in the targeting step based on their squared or cubed form. In the financial application, since the number of initial predictors is smaller, we include all interactions, as well as squared and cubed transformations in the targeting step and feed all predictors entering any of the transformations selected in the targeting step to the prediction step in their original form.[‡]

Table 2 reports the results in the same format as Table 1. Our main conclusions from the former section carry over to the case of targeting with nonlinearities. In addition, there is some improvement in the performance of TRF relative to RF at the one-month forecasting horizon. This suggests that nonlinearities are particularly important at the $h = 1$ horizon and that targeting of predictors in some cases only reaches its full potential if extended to accommodate interactions and nonlinearities among predictors.

---

[‡]We also examined the macroeconomic applications with all interactions and squared terms included in the targeting step. This amounts to more than 5,000 predictors, which is not infeasible, but overspecified. Results are consistent with those reported, and available upon request.

## 4. Concluding remarks

Although RF is applicable in high-dimensional settings due to its potential to detect informative predictors automatically, recent literature highlights the need to target a subset of predictors prior to the final prediction or forecasting step in a given application. The overall prediction from an RF comes about by averaging across the individual predictions, i.e., the trees in the forest. The strength of an individual tree, in an MSE sense, depends on the probability of splitting along strong predictors. Averaging across trees serves to reduce variance. Thus, the benefits from averaging are reduced if trees are similar.

We show in this paper that an initial targeting step added to the RF algorithm serves as an important complement to feature sampling, enabling control of the lower bound on the probability of splitting on strong predictors. We further show that there is an immediate and certain gain in tree strength from (proper) targeting and quantify this. On the other hand, TRF (i.e., RF with an initial targeting step) cannot be expected to perform uniformly better than ordinary RF, since the targeting step likely induces increased correlation across individual trees in the forest, thus reducing the benefit from averaging. This leads to a bias-variance trade-off—in particular, a tree strength-correlation trade-off.

Our empirical analysis covering classical macroeconomic and financial areas of application examines the magnitude of the impact of targeting as well as this trade-off. Based on the empirical applications, we conclude that targeting is useful, particularly if a medium-sized set of predictors is targeted consisting of the best 10–30% of standard initial predictors, since this essentially balances the trade-off between strength and diversity across trees. A medium degree of targeting provides a safety net, ensuring that performance is not significantly reduced and often yielding considerable improvements. In our applications, the gains in predictive accuracy of TRF relative to ordinary RF are substantial in many cases, up to 12–13%, and arise particularly at long forecast horizons, presumably cases with weak signals, and both in expansions and recessions.

# Appendix: Proofs

*Proof of Theorem 1.* The upper bound in (5) follows because splits must be placed along directions in $\mathcal{M}_{try}^{\mathcal{A}}$. For the lower bound, it suffices to argue that if $2\delta_n(\mathcal{A}) < C^\star(\mathcal{A})$, then the split will be placed along a direction in $\mathcal{S}$, i.e.,

$$\sup_{i \in \mathcal{M}_{try}^{\mathcal{A}}, \tau \in A^{(i)}} L_n(i,\tau) > \sup_{i \in \mathcal{A} \setminus \mathcal{S}, \tau \in A^{(i)}} L_n(i,\tau). \tag{A.1}$$

By the definition of $\delta_n(\mathcal{A})$, we have $L_n(i,\tau) \geq L^\star(i,\tau) - \delta_n(\mathcal{A})$, and hence

$$\sup_{i \in \mathcal{M}_{try}^{\mathcal{A}}, \tau \in A^{(i)}} L_n(i,\tau) \geq C^\star(\mathcal{A}) - \delta_n(\mathcal{A}). \tag{A.2}$$

Since $L^\star(i,\tau) = 0$ for $i \notin \mathcal{S}$ (cf. (9)),

$$\sup_{i \in \mathcal{A} \setminus \mathcal{S}, \tau \in A^{(i)}} L_n(i,\tau) = \sup_{i \in \mathcal{A} \setminus \mathcal{S}, \tau \in A^{(i)}} |L_n(i,\tau) - L^\star(i,\tau)| \leq \delta_n(\mathcal{A}). \tag{A.3}$$

Combination of (A.2) and (A.3) shows that

$$\sup_{i \in \mathcal{M}_{try}^{\mathcal{A}}, \tau \in A^{(i)}} L_n(i,\tau) \geq C^\star(\mathcal{A}) - \delta_n(\mathcal{A}) \geq \sup_{i \in \mathcal{A} \setminus \mathcal{S}, \tau \in A^{(i)}} L_n(i,\tau) + (C^\star(\mathcal{A}) - 2\delta_n(\mathcal{A})).$$

Thus, (A.1) holds for $2\delta_n(\mathcal{A}) < C^\star(\mathcal{A})$, hence concluding the proof of the first statement of the Theorem.

For the second statement, assume for simplicity (but without loss of generality) that $a = 1$ and $A = [0,1]$. Then, suppressing dependence on the direction index $i$, it suffices to show that, almost surely,

$$\frac{1}{n} \sum_{k=1}^{n} (Y_k - \bar{Y}_n(\tau))^2 \mathbb{1}_{\{X_k \leq \tau\}} \longrightarrow \mathbb{E}[(Y - \mathbb{E}[Y \mid X \leq \tau])^2 \mathbb{1}_{\{X \leq \tau\}}], \tag{A.4}$$

uniformly in $\tau \in [0,1]$, with $\bar{Y}_n(\tau) = N_n(\tau)^{-1} \sum_{k=1}^{n} Y_k \mathbb{1}_{\{X_k \leq \tau\}}$ for $N_n(\tau) := \sum_{k=1}^{n} \mathbb{1}_{\{X_k \leq \tau\}} \neq 0$ and $\bar{Y}_n(\tau) = 0$ otherwise. By the decomposition

$$\sum_{k=1}^{n} (Y_k - \bar{Y}_n(\tau))^2 \mathbb{1}_{\{X_k \leq \tau\}} = \sum_{k=1}^{n} Y_k^2 \mathbb{1}_{\{X_k \leq \tau\}} - \bar{Y}_n(\tau)^2 N_n(\tau), \tag{A.5}$$

and the fact that

$$\frac{1}{n} \sum_{k=1}^{n} Y_k^2 \mathbb{1}_{\{X_k \leq \tau\}} \longrightarrow \mathbb{E}[Y^2 \mathbb{1}_{\{X \leq \tau\}}], \tag{A.6}$$

uniformly in $\tau$, it suffices to show that

$$\frac{\bar{Y}_n(\tau)^2 N_n(\tau)}{n} \longrightarrow \frac{\mathbb{E}[Y \mathbb{1}_{\{X \leq \tau\}}]^2}{\mathbb{P}(X \leq \tau)}, \tag{A.7}$$

uniformly. Uniform convergence in (A.6) follows directly from the law of large numbers (LLN) for random variables taking values in the space $D([0,1])$ of càdlàg functions (see

Rao, 1963, Theorem 1). Write $\bar{Y}_n(\tau)^2 N_n(\tau)/n = a_n(\tau)^2/b_n(\tau)$, with $a_n(\tau) = \frac{1}{n}\sum_{k=1}^n Y_k \mathbb{1}_{\{X_k \leq \tau\}}$ and $b_n(\tau) = \frac{1}{n} N_n(\tau)$. The LLN for $D([0,1])$-valued random variables implies again that $a_n(\tau) \to a(\tau) := \mathbb{E}[Y\mathbb{1}_{\{X \leq \tau\}}]$ and $b_n(\tau) \to b(\tau) := \mathbb{P}(X \leq \tau)$ uniformly. The issue in immediately concluding that (A.7) holds uniformly is that $h: (x,y) \mapsto \mathbb{1}_{\{y \neq 0\}} x^2/y$ is not uniformly continuous on compacts containing points of the form $(x,0)$. To circumvent this, introduce for each $\delta > 0$ the function $h_\delta: (x,y) \mapsto x^2/(y \vee \delta)$. With $\|\cdot\|_\infty$ denoting the uniform norm, we have

$$\|h(a_n, b_n) - h(a,b)\|_\infty \leq \|h(a_n, b_n) - h_\delta(a_n, b_n)\|_\infty + \|h_\delta(a_n, b_n) - h_\delta(a,b)\|_\infty \\ + \|h_\delta(a,b) - h(a,b)\|_\infty. \tag{A.8}$$

We need to show that the right-hand side converges to zero as $n \to \infty$ with probability one. By Hölder's inequality,

$$\|h_\delta(a,b) - h(a,b)\|_\infty \leq 2 \sup_{\tau: 0 < b(\tau) \leq \delta} \frac{\mathbb{E}[|Y|\mathbb{1}_{\{X \leq \tau\}}]^2}{b(\tau)} \leq 2\mathbb{E}[|Y|^\gamma]^{2/\gamma} \delta^{1-2/\gamma},$$

and

$$\|h(a_n, b_n) - h_\delta(a_n, b_n)\|_\infty \leq 2 \sup_{\tau: 0 < b_n(\tau) \leq \delta} \frac{\left(\frac{1}{n}\sum_{k=1}^n Y_k \mathbb{1}_{\{X_i \leq \tau\}}\right)^2}{b_n(\tau)} \leq 2C^{2/\gamma} \delta^{1-2/\gamma},$$

with $C = \sup_n n^{-1} \sum_{k=1}^n |Y_k|^\gamma$. Fix $\omega$ belonging to the event

$$\left\{\sup_n n^{-1} \sum_{k=1}^n |Y_k|^\gamma < \infty\right\} \cap \left(\bigcap_{\delta \in \mathbb{Q}_+} \{\|h_\delta(a_n, b_n) - h_\delta(a,b)\|_\infty \to 0\}\right),$$

which by Assumption (A1) and $\mathbb{E}[|Y|^\gamma] < \infty$ has probability one. For given $\varepsilon > 0$, we can choose $\delta \in \mathbb{Q}_+$ such that $2\delta^{1-2/\gamma}(C(\omega) \vee \mathbb{E}[|Y|^\gamma])^{2/\gamma} \leq \varepsilon/3$. Moreover, there exists $N = N(\omega) \geq 1$ such that $\|h_\delta(a_n, b_n) - h_\delta(a,b)\|_\infty \leq \varepsilon/3$. Hence, $\|h(a_n, b_n) - h(a,b)\|_\infty \leq \varepsilon$ for all $n \geq N$ by (A.8). Thus, (A.4) holds, and $\delta_n(\mathscr{A}) \to 0$ with probability one.

If (A3) holds, then (9) shows that $C^\star(\mathscr{A}) > 0$ if and only if $\mathscr{M}^{\mathscr{A}}_{try} \cap \mathscr{S} \neq \emptyset$ and, thus, $\mathbb{1}_{\{2\delta_n(\mathscr{A}) < C^\star(\mathscr{A})\}} \to \mathbb{1}_{\{\mathscr{M}^{\mathscr{A}}_{try} \cap \mathscr{S} \neq \emptyset\}}$ almost surely. The convergence $\mathbb{P}(2\delta_n(\mathscr{A}) < C^\star(\mathscr{A})) \to \mathbb{P}(\mathscr{M}^{\mathscr{A}}_{try} \cap \mathscr{S} \neq \emptyset)$ follows by Lebesgue's theorem (Billingsley, 1999, Theorem 16.4), showing that the length of the interval (5) asymptotically shrinks to zero and effectively pushes $\rho_n(\mathscr{A})$ towards its upper bound. □

*Proof of Proposition 1.* Let $\bar{F}_{\mathscr{A}}(x) = \mathbb{P}(C^\star(\mathscr{A}) > x)$ and $\bar{F}_{\mathscr{B}}(x) = \mathbb{P}(C^\star(\mathscr{B}) > x)$ denote the tails of $C^\star(\mathscr{A})$ and $C^\star(\mathscr{B})$, respectively. By the assumptions on stochastic ordering, $\bar{F}_{\mathscr{A}} \leq \bar{F}_{\mathscr{B}}$, and $\mathbb{E}[g(\delta_n(\mathscr{A}))] \leq \mathbb{E}[g(\delta_n(\mathscr{B}))]$ for any non-increasing function $g: [0, \infty) \to [0, \infty)$. Thus, as $\delta_n(\mathscr{A})$ and $C^\star(\mathscr{A})$ are independent, and so are $\delta_n(\mathscr{B})$ and $C^\star(\mathscr{B})$, it follows by the law of total probability that

$$\mathbb{P}(2\delta_n(\mathscr{A}) > C^\star(\mathscr{A})) = \mathbb{E}[\bar{F}_{\mathscr{A}}(2\delta_n(\mathscr{A}))] \leq \mathbb{E}[\bar{F}_{\mathscr{B}}(2\delta_n(\mathscr{B}))] = \mathbb{P}(2\delta_n(\mathscr{B}) > C^\star(\mathscr{B})).$$

This completes the proof. □

*Proof of Proposition 2.* Since only the first direction is strong, $C^\star = \sup_\tau L^\star(1,\tau)$. If $U_\gamma$ is uniform on $[0,\gamma]$, then
$$\text{Var}(\sin(\alpha U_\gamma)) = \frac{1}{2} - \frac{\sin(2\alpha\gamma)}{4\alpha\gamma} - \left(\frac{1-\cos(\alpha\gamma)}{\alpha\gamma}\right)^2.$$
Using this fact and that $\sin(2\alpha(1-\tau)) = -\sin(2\alpha\tau)$ and $\cos(2\alpha(1-\tau)) = \cos(2\alpha\tau)$, it follows that $L^\star(1,\tau) = 2(1-\cos(\alpha\tau))^2/(\alpha^2\tau(1-\tau))$. By the mean value theorem, it follows that
$$L^\star(1,\tau) \le 4\min\left\{\tau, 1-\tau, \frac{1}{\alpha^2\tau(1-\tau)}\right\}. \tag{A.9}$$
By symmetry, take $\tau \le 1/2$. If $\tau(1-\tau) \le \alpha^{-1}$, then $1-\tau \ge 1-2\alpha^{-1}$, so
$$\tau \le \frac{1}{\alpha(1-\tau)} \le (\alpha-2)^{-1}.$$
By (A.9), this shows that $L^\star(1,\tau) \le 4(\alpha-2)^{-1}$. If, on the other hand, $\alpha\tau(1-\tau) \ge 1$, it follows directly from (A.9) that
$$L^\star(1,\tau) \le \frac{4}{\alpha^2\tau(1-\tau)} \le 4\alpha^{-1},$$
thus concluding the proof. □

*Proof of Theorem 2.* Given an interval $A \subseteq [0,1]$, it can be verified, using Assumptions (A2)–(A3), that (4) is optimized at its midpoint $\tau^\star$, with
$$L^\star(\tau^\star, A) = \beta_1^2 \frac{\text{Leb}(A)^2}{16}. \tag{A.10}$$
Under the splitting rule (R), this implies that the targeted tree is grown by first splitting $[0,1]$ (level 0) into $[0,1/2]$ and $[1/2,1]$ (level 1), then splitting all intervals in level 1 before splitting at the next level, and so on. In particular, $\text{MSE}(\widetilde{f}_{L+1}) = \text{MSE}(\widetilde{f}_L) - \beta_1^2 8^{-(k-1)}/16$, with $k \ge 1$ the smallest integer such that $L \le 2^k - 1$. This formula can be used inductively to establish that
$$\text{MSE}(\widetilde{f}_L) = \beta_1^2 \frac{7 \times 2^k - 3L}{48 \times 8^k},$$
with $k \ge 1$ the largest integer such that $L \ge 2^k$, i.e., $k = \lfloor \log_2(L) \rfloor$. This establishes (21). To obtain the bounds (22), we make some initial observations:

**Observation I** Let $A \subseteq [0,1]^p$ be a leaf in the non-targeted tree $\widehat{f}_L$. Then $\{x^{(1)} : x \in A\} = B_i^k$ for some $k = 1,\ldots,L$, with $B_i^k$ a leaf of the targeted tree $\widetilde{f}_k$. This follows since $X^{(1)}$ and $X^{(-1)} := (X^{(2)},\ldots X^{(p)})'$ are independent, by Assumption (A2), and hence $L^\star(i,\tau,A)$ depends only on $A$ through $\{x^{(1)} : x \in A\}$.

iii

**Observation II** Let $(B_i^k)_{i=1}^k$ be the partition of $[0,1]$ associated with the $k$th level targeted tree $\widetilde{f}_k$, and consider a leaf $B = B_i^k$. Then

$$\text{MSE}(\widetilde{f}_{L+1} \mid X^{(1)} \in B) \leq \text{MSE}(\widetilde{f}_L \mid X^{(1)} \in B), \qquad \text{for any } L \geq k, \tag{A.11}$$

with equality if and only if the partitions $(B_j^L \cap B)_{j=1}^L$ and $(B_j^{L+1} \cap B)_{j=1}^{L+1}$ are identical. Equality in (A.11) follows if the aforementioned partitions are the same, since in this case $\widetilde{f}_{L+1}(x) = \widetilde{f}_L(x)$ for all $x = (x^{(1)}, x^{(-1)})$ with $x^{(1)} \in B$. If they are not the same, the $L$th split $\tau^\star$ is performed in a subset $\bar{B} \in \{B_j^L : j = 1, \ldots, L\}$ of $B$, so

$$\mathbb{E}_{\bar{B}}[(Y - \widetilde{f}_{L+1}(X))^2] = \mathbb{E}_{\bar{B}}[(Y - \mathbb{E}_{\bar{B}}[Y])^2 \mathbb{1}_{\{X^{(1)} \leq \tau^\star\}}] + \mathbb{E}_{\bar{B}}[(Y - \mathbb{E}_{\bar{B}}[Y])^2 \mathbb{1}_{\{X^{(1)} > \tau^\star\}}]$$
$$= \mathbb{E}_{\bar{B}}[(Y - \widetilde{f}_L(X))^2] - L^\star(\tau^\star, \bar{B}),$$

subscript $\bar{B}$ indicating conditioning on the event $\{X^{(1)} \in \bar{B}\}$. As already noted, $L^\star(\tau^\star, \bar{B}) = \beta_1^2 \text{Leb}(\bar{B})^2/16 > 0$, and since $\widetilde{f}_L(x)$ and $\widetilde{f}_{L+1}(x)$ are identical for $x = (x^{(1)}, x^{(-1)})$ with $x^{(1)} \in B \setminus \bar{B}$ and $\mathbb{P}(X^{(1)} \in \bar{B}) > 0$, the inequality in (A.11) will be strict.

To obtain the lower bound in (22), note that the first $\ell_1 - 1$ splits have been weak, and denote by $T_1, \ldots, T_{\ell_1}$ the associated $\ell_1$ subtrees. We will argue that if $N \leq \ell_1(2^m - 1)$ for a given $m \geq 1$, then the strength of a given subtree $T$ is no better than that of $\widetilde{f}_{2^m}$. Initially, note that the construction of $\widetilde{f}_{2^m}$ implies that $\text{Leb}(B_j^{2^m}) = 2^{-m}$, so we can fix $j$ such that

$$\text{Leb}(B_j^{2^m-1}) > 2^{-m}. \tag{A.12}$$

We proceed to proof by contradiction. Thus, suppose $A = A^\mathscr{S} \times A^\mathscr{W}$ is a leaf of $T$ with

$$\text{MSE}_\Theta(T \mid X \in A) < \text{MSE}_\Theta(\widetilde{f}_{2^m} \mid X \in A). \tag{A.13}$$

By Observation I, there exists $k$ such that $A^\mathscr{S} = B_i^{k+1}$, for some $i$. Moreover, due to (A.11) and (A.13), it must be the case that $k > 2^m - 1$, and that $A^\mathscr{S}$ has been obtained by performing splits in one of the sets $(B_i^{2^m})_{i=1}^{2^m}$. At the same time, since we have at most $\ell_1(2^m - 1)$ strong splits in total, this means that at least one of the other subtrees $\bar{T}$ has received strictly less than $2^m - 1$ strong splits. In particular, by (A.12), there exists a leaf $\bar{A} = \bar{A}^\mathscr{S} \times \bar{A}^\mathscr{W}$ in $\bar{T}$ such that $\text{Leb}(\bar{A}^\mathscr{S}) > 2^{-m}$. This is a contradiction, since (R) together with (A.10) imply that no split will be placed in $(B_i^{2^m})_{i=1}^{2^m}$ before $\bar{A}^\mathscr{S}$ has been split. Since this analysis holds for all subtrees $T_1, \ldots, T_{\ell_1}$, we conclude that

$$\text{MSE}_\Theta(\widehat{f}_L) \geq \text{MSE}_\Theta(\widetilde{f}_{2^m}), \qquad \text{if } N \leq \ell_1(2^m - 1).$$

Thus, we may choose $m = \lceil \log_2(1 + N/\ell_1) \rceil$, which yields the result.

For the upper bound in (22), note that after the first $\ell_0 - 1$ splits, the non-targeted tree is identical to $\widetilde{f}_{\ell_0}$. If $m \geq 1$ is an integer such that $2^m \geq \ell_0$, then, given that no weak splits occur, it takes $2^m - \ell_0$ strong splits before any given leaf $A = A^{\mathscr{S}} \times A^{\mathscr{W}}$ satisfies $\text{Leb}(A^{\mathscr{S}}) = 2^{-m}$. While we do indeed have $L - N - 1$ weak splits, none of the first $\ell_0$ nodes can possibly have been divided into more than $L - N$ branches by weak splits. Consequently, as long as we still have at least $\ell_0(2^m - \ell_0)(L - N)$ strong splits available, any leaf $A$ must satisfy $\text{Leb}(A^{\mathscr{S}}) \leq 2^{-m}$. At the same time, we cannot have $\text{MSE}_\Theta(\widehat{f}_L \mid X \in A) > \text{MSE}_\Theta(\widetilde{f}_{2^m} \mid X \in A)$, since this would imply $A^{\mathscr{S}} = B_i^k$ for some $k < 2^m$ and, hence, $\text{Leb}(A^{\mathscr{S}}) > 2^{-m}$. This allows concluding that

$$\text{MSE}_\Theta(\widehat{f}_L) \leq \text{MSE}_\Theta(\widetilde{f}_{2^m}), \qquad \text{if } N - \ell_0 + 1 \geq \ell_0(2^m - \ell_0)(L - N).$$

In particular, we can set $m = \lfloor \log_2(\ell_0 + \frac{N - \ell_0 + 1}{\ell_0(L - N)}) \rfloor$, which verifies the upper bound in (22) and concludes the proof.

□

*Proof of Corollary 1.* From Theorem 2 it follows that

$$g_0(x, y) \geq \mathbb{E}\big[(f(X) - \widehat{f}_L(X))^2 \mid N = x, \ell_0 = y, \ell_1 = z\big] \geq g_1(x, z), \tag{A.14}$$

so the law of total expectation implies (23). To derive the distribution of $(N, \ell_0)$, define i.i.d. Bernoulli random variables by $Z_k = \mathbb{1}_{\{\mathscr{M}_{try}^{(k)} \cap \mathscr{S} \neq \emptyset\}}$ for $k = 1, \ldots, L - 1$, so that $\rho = \mathbb{P}(Z_k = 1) = 1 - \mathbb{P}(Z_k = 0)$. We always have $N = n$ for some integer $n \in [1, L - 1]$, so we only have to consider such $n$. Let $n \in (0, L - 1)$, and note that both weak and strong splits have occured when $N = n$. Moreover, it must necessarily be the case that $\ell_0 = k$, for some integer $k \in [1, n + 1]$. For any such pair $(n, k)$, we thus find that

$$\mathbb{P}(N = n, \ell_0 = k) = \mathbb{P}\Big(\sum_{m=1}^{L-1} Z_m = n, Z_1 = \cdots = Z_{k-1} = 1, Z_k = 0\Big)$$
$$= \rho^{k-1}(1 - \rho) \text{Bin}(n + 1 - k; L - 1 - k),$$

using the fact that $\mathbb{P}(\sum_{k \in A} Z_k = n) = \text{Bin}(n; |A|, \rho)$ for an arbitrary set $A \subseteq \{1, \ldots, L - 1\}$. If $N = 0$, then all splits have been weak, and, necessarily, $\ell_0 = 1$. This means that

$$\mathbb{P}(N = 0, \ell_0 = 1) = \mathbb{P}(Z_1 = \cdots = Z_{L-1} = 0) = (1 - \rho)^{L-1}.$$

On the other hand, if $N = L - 1$, then all splits have been strong, and $\ell_0 = 1$ by convention, so

$$\mathbb{P}(N = L - 1, \ell_0 = 1) = \mathbb{P}(Z_1 = \cdots = Z_{L-1} = 1) = \rho^{L-1}.$$

Combination of these findings verifies the form of $\mathbb{P}(N = n, \ell_0 = k)$ stated in the corollary. Similar considerations verify the form of $\mathbb{P}(N = n, \ell_1 = k)$.

□